\documentclass[twocolumn, showpacs, nofootinbib, aps, prd]{revtex4}

\usepackage{graphicx}
\usepackage{dcolumn}
\usepackage{color}
\usepackage{amsfonts}
\usepackage[colorlinks=true,linkcolor=blue,citecolor=blue, urlcolor=blue]{hyperref}

\begin{document}

\title{Heavy Pseudoscalar Leading-Twist Distribution Amplitudes within QCD Theory in Background Fields}

\author{Tao Zhong$^{1}$}
\email{zhongtao@ihep.ac.cn}
\author{Xing-Gang Wu$^{2}$}
\email{wuxg@cqu.edu.cn}
\author{Tao Huang$^1$}
\email{huangtao@ihep.ac.cn}

\address{$^1$ Institute of High Energy Physics and Theoretical Physics Center for Science Facilities, Chinese Academy of Sciences, Beijing 100049, P.R. China \\
$^{2}$ Department of Physics, Chongqing University, Chongqing 401331, P.R. China}

\date{\today}

\begin{abstract}
In this paper, we study the leading-twist distribution amplitude (DA) of the heavy pseudoscalars (HPs), such as $\eta_c$, $\eta_b$ and $B_c$, within the QCD theory in the background fields. New sum rules up to dimension-six condensates for both the HP decay constants and their leading-twist DA moments are presented. From the sum rules for the HP decay constants, we obtain $f_{\eta_c} = 453 \pm 4 \textrm{MeV}$, $f_{B_c} = 498 \pm 14 \textrm{MeV}$, and $f_{\eta_b} = 811 \pm 34 \textrm{MeV}$. Basing on the sum rules for the HPs' leading-twist DA moments, we construct a new model for the $\eta_c$, $\eta_b$ and $B_c$ leading-twist DAs. Our present HP DA model can also be adaptable for the light pseudo-scalar DAs, such as the pion and kaon DAs. Thus, it shall be applicable for a wide range of QCD exclusive processes. As an application, we apply the $\eta_c$ leading-twist DA to calculate the $B_c \to \eta_c$ transition form factor $f_+^{B_c \to \eta_c}(q^2)$. At the maximum recoil region, we obtain $f_+^{B_c \to \eta_c}(0) = 0.612^{+0.053}_{-0.052}$. After further extrapolating the TFF $f_+^{B_c \to \eta_c}(q^2)$ to its allowable $q^2$ region, we predict the branching ratio for the semi-leptonic decay $B_c \to \eta_c l \nu$. We obtain ${\cal B}(B_c \to \eta_c l \nu)=\left(7.70^{+1.65}_{-1.48}\right) \times 10^{-3}$ for massless leptons, which is consistent with the LCSRs estimation obtained in the literature.
\end{abstract}

\pacs{12.38.-t, 12.38.Bx, 14.40.Aq}

\maketitle

\section{introduction}

The hard exclusive processes involving the heavy pseudo-scalars (HPs), such as $\eta_c$, $\eta_b$ and $B_c$, have been studied within several approaches, such as the perturbative QCD (pQCD) factorization approach~\cite{HP_PQCD12,HP_PQCD13,TFF13,HP_PQCD14,chang}, the non-relativistic QCD (NRQCD) factorization approach~\cite{BRNRQCD13,qiao12,shen}, and the QCD light-cone sum rules (LCSRs) approach~\cite{HP_LCSR,TFF,BRSR08}. The HP leading-twist distribution amplitude (DA) is always an important input for those analysis, and a more precise DA shall lead to more precise prediction.

The HP leading-twist DA at the scale $\mu$ can be expanded in Gegenbauer polynomials as~\cite{HPDA_CZ}:
\begin{eqnarray}
\phi_{\rm HP}(\mu,x) = 6x(1-x) \left[ 1 + \sum_{n=1}^\infty a_n^{\rm HP}(\mu) C^{3/2}_n(2x-1) \right],  \label{HPDA_CZ}
\end{eqnarray}
where $a_n^{\rm HP}(\mu)$ stands for the $n_{\rm th}$-order Gegenbauer moment, and the odd moments should be zero for the $\eta_c$ and $\eta_b$ mesons. When the scale $\mu$ tends to infinity, the DA $\phi_{\rm HP}(\mu,x)$ shall evolve into its asymptotic form $6x(1-x)$~\cite{EE}. Since the typical energy scale of a specific process is always finite, it is interesting to know the $\phi_{\rm HP}$ behavior at any finite scale.

It is reasonable to assume that the $\eta_c$ and $\eta_b$ DAs have similar behaviors. As for the $\eta_c$ leading-twist DA, several models have been suggested in the literature~\cite{HP_LCSR,ECDA_BC,ECDA_BLL,ECDA_BHL,ECDA_BHL1, ECDA_BKL,ECDA_MS,ECDA_CWH}. For examples, Bondar and Chernyak~\cite{ECDA_BC} proposed a phenomenological model for the $\eta_c$ leading-twist DA (the BC model) as a try to resolve the disagreement between the experimental observations and the NRQCD prediction on the production cross section of $e^+ e^- \to J/\Psi+ \eta_c$; Braguta, Likhoded and Luchinsky~\cite{ECDA_BLL} proposed a model for the $\eta_c$ leading-twist DA (the BLL model) based on the moments calculated under the QCD Shifman-Vainshtein-Zakharov (SVZ) sum rules up to dimension-four condensates. As for the $B_c$ meson, one usually adopts a naive $\delta$-like model for its leading-twist DA $\phi_{B_c}$~\cite{BCDA}.

In this paper, we study the HP leading-twist DAs within the SVZ sum rules~\cite{SVZ} under the background field theory (BFT)~\cite{BFT}. As the basic assumption of the SVZ sum rules, the quark condensate $\left<\bar{q}q\right>$, the gluon condensate $\left<G^2\right>$ and etc., reflect the nonperturbative property in QCD. It is noted that the BFT provides a self-consistent description for those vacuum condensates and provides a systematic way to achieve the goal of the SVZ sum rules~\cite{BFT}. The HP DAs are more involved than the light pseudoscalar DAs, since we have to take the quark mass effect in the calculation. Recently, within the framework of BFT, we have for the first time calculated the quark propagator and vertex operator $(z\cdot\tensor{D})^n$ with full mass dependence up to dimension-six operators~\cite{BFSR}. Thus we are facing the chance of deriving a more precise sum rules for the HP DA moments and a precise HP DA behavior. For convenience, based on the BHL-prescription for constructing the meson
wavefunctions~\cite{BHL}, we suggest a general model for the HP leading-twist wavefunctions and their DAs.

As an application of the suggested DA model, we apply the $\eta_c$ leading-twist DA to calculate the $B_c\to \eta_c$ transition form factor (TFF) $f_+^{B_c \to \eta_c}(q^2)$ within the LCSRs. It is the key component for the semi-leptonic decay $B_c \to \eta_c l \nu$. It is also the only TFF for the decay if the generated leptons are massless. By adopting the conventional correlator for the LCSRs, similar to the $B\to\pi$ TFFs~\cite{ball}, the TFF $f_+^{B_c \to \eta_c}(q^2)$ shall be formulated as the function involving the
$\eta_c$ leading-twist DA, twist-3 DA, and other higher-twist DAs. The higher-twist DAs follow the power suppression rule in large scale region, however they may have sizable contributions to the TFF in the intermediate energy regions, similar to the pionic cases of the $B\to\pi$ TFFs and the pion TFFs~\cite{wuhuang}. At present, the $\eta_c$ higher-twist DAs are still with great uncertainty, thus the possible LCSRs with $\eta_c$ various twist DAs shall inversely greatly dilute our understanding of the leading-twist DA behaviors. To cure the problem, we adopt the chiral correlator suggested in
Ref.~\cite{TFF} to do our calculation, and we find that the most uncertain twist-3 DAs can be eliminated, then we can see more clearly on how the leading-twist DA affects $f_+^{B_c \to \eta_c}(q^2)$.

The remaining parts of the paper are organized as follows. In Sec.II, the QCD SVZ sum rules for the HP decay constants and the HP leading-twist DA moments are given within the framework of BFT. A new model for the HP leading-twist DAs are also suggested here. Numerical results are presented in Sec.III. Sec.IV is reserved for a summary.

\section{calculation technology}

\subsection{SVZ Sum Rules for the HP Decay Constants}

To obtain the SVZ sum rules for the HP decay constants, we take the following correlation function
\begin{eqnarray}
\Pi(q^2) = i \int d^4x e^{iq\cdot x} \left<0\left|\textrm{T} \left\{J_5(x) J_5^\dagger(0)\right\} \right|0\right>. \label{fcorrelator}
\end{eqnarray}
Here the pseudo-scalar current
\begin{eqnarray}
J_5(x) = \bar{Q}_1(x) i\gamma_5 Q_2(x), \label{J5}
\end{eqnarray}
where $Q_{1}=b$ and $Q_{2}=c$ for $B_c$, $Q_{1}=Q_{2}=c$ ($Q_{1}=Q_{2}=b$) for $\eta_{c}$ ($\eta_{b}$), respectively. The HP decay constant $f_{\rm HP}$ is defined as
\begin{eqnarray}
\left<0\left|J_5\right|HP\right> = f_{\rm HP} \frac{m_{\rm HP}^2}{m_1 + m_2}, \label{f_definition}
\end{eqnarray}
where $m_{\rm HP}$ stands for the HP mass and $m_{1(2)}$ is the mass of $Q_{1(2)}$ quark.

Following the standard sum rules procedures, the correlation function (\ref{fcorrelator}) can be inserted by a completed set of intermediate hadronic states in the physical region. It can also be treated in the framework of the operator product expansion (OPE) in the deep Euclidean region simultaneously. Those two results can be related by the dispersion relation
\begin{eqnarray}
\Pi_{\rm QCD}(q^2) = \frac{1}{\pi} \int^\infty_{t_{\rm min}} ds \frac{\textrm{Im} \Pi_{\rm had}(s)}{s-q^2} + \textrm{subtractions},
\label{disrel_DC}
\end{eqnarray}
where $t_{\rm min} = (m_1 + m_2)^2$. Then, the sum rules can be achieved by applying the Borel transform for both side of Eq.(\ref{disrel_DC}).

\begin{figure*}
\includegraphics[width=0.9\textwidth]{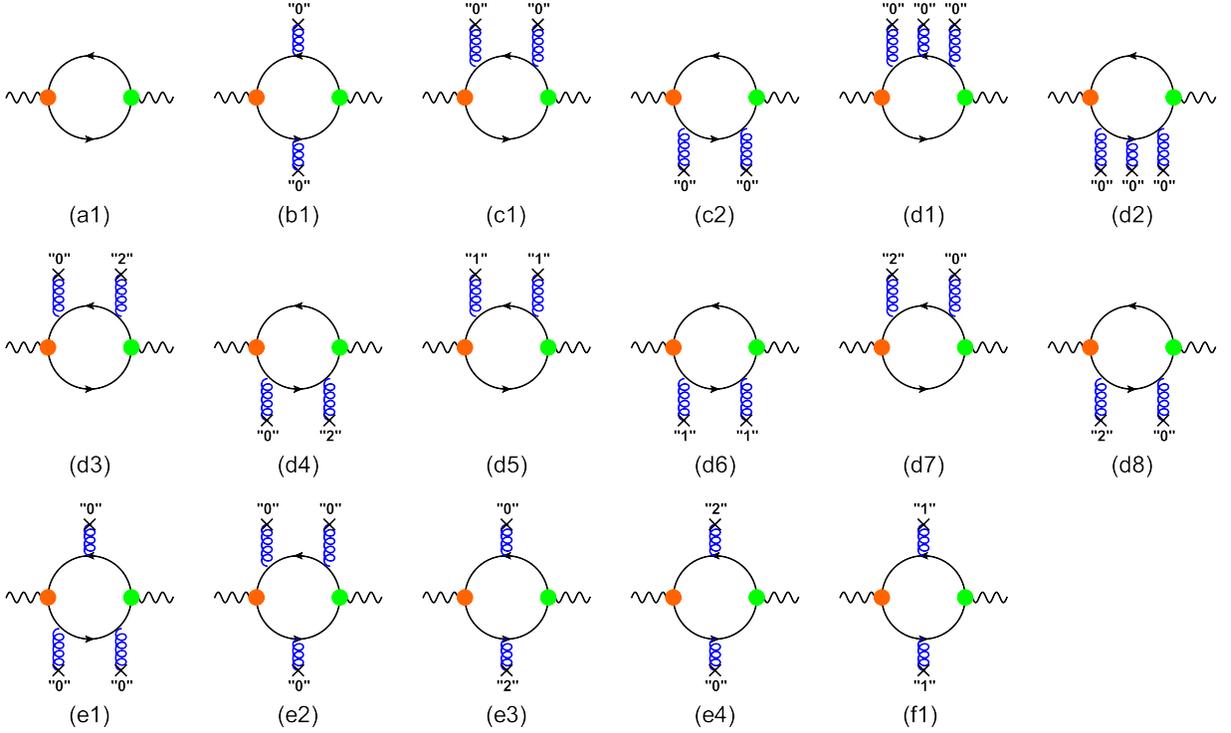}
\caption{Feynman diagrams for the HP decay constant. The big dot stands for the vertex operators $i \gamma_5$ in the current (\ref{J5}), the cross symbol attached to the gluon line indicates the tensor of the local gluon background field, and ``$n$" indicates the $n_{\rm th}$-order covariant derivative. }
\label{feynDC}
\end{figure*}

More explicitly, on the one hand, we do the OPE for the correlator
(\ref{fcorrelator}), $\Pi_{\rm QCD}(q^2)$, within the framework of
BFT. For the purpose, we first apply the following replacement for
the quark fields
\begin{equation}
Q_{1(2)} \to Q_{1(2)} + \eta_{1(2)}   \label{BFT_replace}
\end{equation}
in Eq.(\ref{fcorrelator}), where $Q_{1}$ and $Q_{2}$ in the right-hand-side of Eq.(\ref{BFT_replace}) stand for the quark background fields, $\eta_{1}$ and $\eta_{2}$ are the corresponding quantum fluctuations (quantum fields) on the background field. The quantum fields interacts with each other according to the Feynman rule of BFT~\cite{BFT}, for example, the quantum quark-anti-quark pair can be contracted as a propagator; while the remaining background fields shall be kept to form the various vacuum matrix elements. Fig.\ref{feynDC} shows the Feynman diagrams for determining the HP decay constant up to dimension-six operators, in which the newly derived quark propagator with up to dimension-six operators has been adopted~\cite{BFSR}. In Fig.\ref{feynDC}, the big dot stands for the vertex operators $i \gamma_5$ in the current (\ref{J5}), the cross symbol attached to the gluon line indicates the tensor of the local gluon background field, and ``$n$" indicates the $n_{\rm th}$-order covariant derivative. Fig.\ref{feynDC}.(a1) provides the perturbative contribution, Fig.\ref{feynDC}.(b1,b2,c2) provide the contributions proportional to the dimension-four condensate $\left<\alpha_s G^2\right>$, and the remaining thirteen diagrams Fig.\ref{feynDC}.(d1-f1) provide the contributions proportional to the dimension-six condensate $\left<g_s^3fG^3\right>$. Here, $\left<\alpha_s G^2\right>$ and $\left< g_s^3 fG^3 \right>$ are abbreviations for the condensates $\left<0\left|\alpha_s G^A_{\mu\nu} G^{A\mu\nu}\right|0\right>$ and $\left<0\left|g_s^3 f^{ABC} G^{A\mu\nu} G^{B\rho}_{\nu} G^{C}_{\rho\mu}\right|0\right>$, respectively, where the color indices $A,B,C = (1,2,\cdots,8)$. Then, we can directly derive the explicit expression for $\Pi_{\rm QCD}(q^2)$ from Fig.\ref{feynDC}, which is rather lengthy and shall not be presented here for simplicity.

On the other hand, with the help of the definition (\ref{f_definition}), the hadronic spectrum representation of the correlator (\ref{fcorrelator}) can be written as
\begin{eqnarray}
\textrm{Im} \Pi_{\rm had}(q^2) &=& \pi \delta (q^2 - m_{\rm HP}^2) \frac{f_{\rm HP}^2 m_{\rm HP}^4}{(m_1+m_2)^2} \nonumber\\
&& + \pi \rho^{\rm cont}(q^2) \theta (q^2 - s_{\rm HP}),
\label{hadron_expression_DC}
\end{eqnarray}
where $s_{\rm HP}$ is the continue threshold parameter, $\theta$ is the usual step function, and $\rho^{\rm cont}$ stands for the hadron spectrum density from the continuous states. Due to the quark-hadron duality, $\rho^{\rm cont}$ can be written as
\begin{eqnarray}
\rho^{\rm cont}(s) = \frac{1}{\pi} \textrm{Im} \Pi_{\rm pert}(s).
\label{density_DC}
\end{eqnarray}

As a combination of Eqs.(\ref{disrel_DC}, \ref{hadron_expression_DC}, \ref{density_DC}), we are ready to derive the SVZ sum rules for the HP decay constant. After further applying the Borel transformation to suppress both the unknown continuous states' contributions and the higher dimensional condensates' contributions, the final SVZ sum rules for the HP decay constant reads
\begin{widetext}
\begin{eqnarray}
\frac{f_{\rm HP}^2 m^4_{\rm HP}}{M^2 (m_1 + m_2)^2} e^{-\left[ m_{\rm HP}^2 / M^2 \right]} = \frac{1}{\pi} \frac{1}{M^2} \int^{s_{\rm HP}}_{t_{\rm min}} ds e^{-s/M^2} \textrm{Im} \Pi_{\rm pert} (s) + \hat{L}_M \Pi_{\left<G^2\right>}(Q^2) + \hat{L}_M \Pi_{\left<G^3\right>}(Q^2), \label{f_srborel}
\end{eqnarray}
\end{widetext}
where $M$ is the Borel parameter, the operator $\hat{L}_M = \lim_{-q^2,n\to\infty;(-q^2/n)=M^2} \frac{(-q^2)^{n}}{(n-1)!} \left(\frac{d}{d q^2}\right)^n$ stands for the usual Borel transformation operator. The perturbative part have been studied up to one-loop level by Ref.~\cite{DCas}, and we have
\begin{widetext}
\begin{eqnarray}
\textrm{Im} \Pi_{\rm pert}(s) &=& \frac{3}{8\pi} \frac{\bar{s}^2}{s} v \left\{ 1 + \frac{4\alpha_s(\mu)}{3\pi v} \left[ (1+v^2) \left( \frac{\pi^2}{6} + \ln \left( \frac{1+v}{1-v} \right) \ln \left( \frac{1+v}{2} \right) + 2\textrm{Li}_2 \left( \frac{1-v}{1+v} \right) + \textrm{Li}_2 \left( \frac{1+v}{2} \right) \right.\right.\right. \nonumber\\
&-& \left. \textrm{Li}_2 \left( \frac{1-v}{2} \right) + \frac{1}{2} \sum_i \left[ -4\textrm{Li}_2 (v_i) + \textrm{Li}_2 (v_i^2) + \textrm{Li}_2 \left( \frac{1+v_i}{2} \right) - \textrm{Li}_2 \left( \frac{1-v_i}{2} \right) \right] \right) \nonumber\\
&+& \left( \frac{19}{16} - 3v + \frac{1}{8}v^2 + \frac{3}{16} v^4 \right) \ln \left( \frac{1+v}{1-v} \right) + \frac{29}{8}v - \frac{3}{8}v^2 + 6v \ln \left( \frac{1+v}{2} \right) - 4v\ln v \nonumber\\
&+& \frac{1}{2}(1+v^2) \sum_i \ln \left( \frac{1+v_i}{1-v_i} \right) \ln \left( \frac{v_i}{v} \right) + \frac{1}{2}v \sum_i \left( \frac{1}{v_i} - \frac{1}{v} \right) \ln \left( \frac{1+v_i}{1-v_i} \right) + v\ln \left( \frac{s}{\bar{s}} \right) + \frac{3}{2}v \ln \left( \frac{m_1}{m_2} \right) \nonumber\\
&+& \left.\left. v \left[ \frac{(m_1 - m_2)^2}{s} v \ln \left( \frac{1+v}{1-v} \right) - \frac{m_2^2 - m_1^2}{s} \ln \left( \frac{m_2}{m_1} \right) \right] \right] \right\},
\end{eqnarray}
\end{widetext}
where $\bar{s} = s - (m_1 - m_2)^2$, $v^2 = 1 - 4m_1m_2/\bar{s}$, $v_{1(2)} = \bar{s}v/s_{1(2)}$ with $s_1 = s - m_1^2 + m_2^2$ and $s_2 = s + m_1^2 - m_2^2$, $\textrm{Li}_2(x) = -\int^x_0 dt \frac{\ln (1-t)}{t}$ is Spence function. Moreover, for the parts proportional to the dimension-four and dimension-six condensates, we have
\begin{widetext}
\begin{eqnarray}
\hat{L}_{M} \Pi_{\left<G^2\right>}(q^2) &=& \frac{\left<\alpha_sG^2\right>}{48\pi} \int^1_0 dx \exp \left[ -\frac{m_1^2x + m_2^2(1-x)}{M^2x(1-x)} \right] \left\{ \left[ -6m_1m_2 - 2\frac{m_1^2x^3 + m_2^2(1-x)^3}{x(1-x)} \right] \frac{1}{M^4 x^2(1-x)^2} \right. \nonumber\\
&+& \left. [m_1x + m_2(1-x)]^2 \frac{m_1^2x^3 + m_2^2(1-x)^3}{M^6x^4(1-x)^4} \right\},  \\
\hat{L}_{M} \Pi_{\left<G^3\right>}(q^2) &=& \frac{\left<g_s^3fG^3\right>}{32(4\pi)^2} \int^1_0 dx \exp \left[ -\frac{m_1^2x + m_2^2(1-x)}{M^2x(1-x)} \right] \left( \left[ 64 - \frac{416}{3}x(1-x) - 45\frac{x^3 + (1-x)^3}{x(1-x)} \right] \right. \nonumber\\
&\times& \frac{1}{M^4x^2(1-x)^2} + \left\{ \left( 45 \frac{[m_1x + m_2(1-x)]^2}{x(1-x)} - \frac{2}{3} m_1m_2 \right) [x^3 + (1-x)^3] - 6\frac{m_1^2x^4 + m_2^2(1-x)^4}{x(1-x)} \right. \nonumber\\
&-& 12[m_1^2x^3 + m_2^2(1-x)^3] - \frac{478}{9} m_1m_2x(1-x) + \frac{287}{9} [m_1x + m_2(1-x)]^2 - \frac{320}{9} x(1-x) \nonumber\\
&\times& \left. [m_1^2x + m_2^2(1-x)] \right\} \frac{1}{2M^6x^3(1-x)^3} + \left\{ \left( 6\frac{[m_1x + m_2(1-x)]^2}{x(1-x)} + 28m_1m_2 \right) [m_1^2x^4 + m_2^2(1-x)^4] \right. \nonumber\\
&-& 4[m_1^4x^4 + m_2^4(1-x)^4] + \frac{16}{5} \frac{m_1^4x^5 + m_2^4(1-x)^5}{x(1-x)} + \left( 30[m_1x + m_2(1-x)]^2 + 88m_1m_2x(1-x) \right) \nonumber\\
&\times& [m_1^2x^2 + m_2^2(1-x)^2] - \frac{112}{3} [m_1x + m_2(1-x)]^4 + \frac{64}{3} m_1m_2x(1-x) [m_1x + m_2(1-x)]^2 \nonumber\\
&+& \left.\left. 128m_1^2m_2^2 x^2(1-x)^2 \right\} \frac{1}{6M^8x^4(1-x)^4} - \frac{2}{15} [m_1x + m_2(1-x)]^2 \frac{m_1^4x^5 + m_2^4(1-x)^5}{M^{10} x^6(1-x)^6} \right).
\end{eqnarray}
\end{widetext}
In deriving these sum rules, we have adopted the $\overline{MS}$-scheme to deal with the infrared divergences. During the calculation, we have to deal with the following vacuum matrix elements in $D$-dimensional space ($D=4-2\epsilon$): $\left<0\left| G^A_{\mu\nu} G^B_{\rho\sigma} \right|0\right>$, $\left<0\left| G^A_{\mu\nu} G^B_{\rho\sigma} G^C_{\lambda\tau} \right|0\right>$, $\left<0\left| G^A_{\mu\nu;\lambda} G^B_{\rho\sigma;\tau} \right|0\right>$, $\left<0\left| G^A_{\mu\nu;\lambda\tau} G^B_{\rho\sigma} \right|0\right>$ and $\left<0\left| G^A_{\mu\nu} G^B_{\rho\sigma;\lambda\tau} \right|0\right>$. The formulae for
relating these matrix elements with the conventional condensates under the $D$-dimensional space have been given in the Appendix B of Ref.~\cite{BFSR}. For simplicity, we do not present them here, and the interesting readers may turn to this reference for detailed technology.

\subsection{SVZ Sum Rules for the Moments of the HP Leading-Twist DA}

The HP leading-twist DA $\phi_{\rm HP}$ is defined as
\begin{eqnarray} &&
\left<0\left| \bar{Q}_1(z) \not\! z \gamma_5 Q_2(-z) \right|HP(q)\right>
\nonumber\\ && \quad\quad\quad
= i (z\cdot q) f_{\rm HP} \int^1_0 du e^{i\xi(z\cdot q)} \phi_{\rm HP}(u),  \label{DA_definition}
\end{eqnarray}
where $\xi = 2u-1$. Expanding the left-hand-side of Eq.(\ref{DA_definition}) near $z=0$ and writing the exponent in right-hand-side of Eq.(\ref{DA_definition}) as power series, we obtain the definition of the DA moments
\begin{eqnarray}
&& \left<0\left| \bar{Q}_1(0) \not\! z \gamma_5 (iz\cdot \tensor{D})^n Q_2(0) \right|HP(q)\right> \nonumber\\
&& \quad\quad\quad\quad\quad\quad\quad
= i(z\cdot q)^{n+1} f_{\rm HP} \left<\xi^n\right>_{\rm HP},
\label{moment_definition}
\end{eqnarray}
where
\begin{eqnarray}
\left<\xi^n\right>_{\rm HP} = \int^1_0 du (2u-1)^n \phi_{\rm HP}(u)
\label{moment_n}
\end{eqnarray}
is the $n_{\rm th}$-order moment of $\phi_{\rm HP}$. The $0_{\rm th}$-order moment
\begin{eqnarray}
\left<\xi^0\right>_{\rm HP} = \int^1_0 du \phi_{\rm HP}(u) = 1
\label{moment_0}
\end{eqnarray}
gives the normalization condition for $\phi_{\rm HP}$. Setting $n=0$ in Eq.(\ref{moment_definition}), one can get
\begin{eqnarray}
\left<0\left| \bar{Q}_1(0) \not\! z \gamma_5 Q_2(0) \right|HP(q)\right> = i(z\cdot q) f_{\rm HP}.   \label{moment_definition_0}
\end{eqnarray}

To derive sum rules for the $\phi_{\rm HP}$ moments, we consider the following correlation function:
\begin{eqnarray} &&
(z\cdot q)^{n+2} I (q^2) \nonumber\\
&=& i \int d^4x e^{iq\cdot x} \left<0\left| T \left\{ J_n(x) J^\dag_0(0) \right\} \right|0\right>,
\label{correlator}
\end{eqnarray}
where $z^2 = 0$, and the two currents
\begin{eqnarray}
J_n(x) &=& \bar{Q}_1(x) \not\! z \gamma_5 (i z\cdot \tensor{D})^n Q_2(x),
\nonumber\\
J^\dagger_0(0) &=& \bar{Q}_2(0) \not\! z \gamma_5 Q_1(0).
\nonumber
\end{eqnarray}

Similar to Sec.II.A, we can deduce the SVZ sum rules for the moments $\left<\xi^n\right>_{\rm HP}$.

\begin{figure*}
\includegraphics[width=0.9\textwidth]{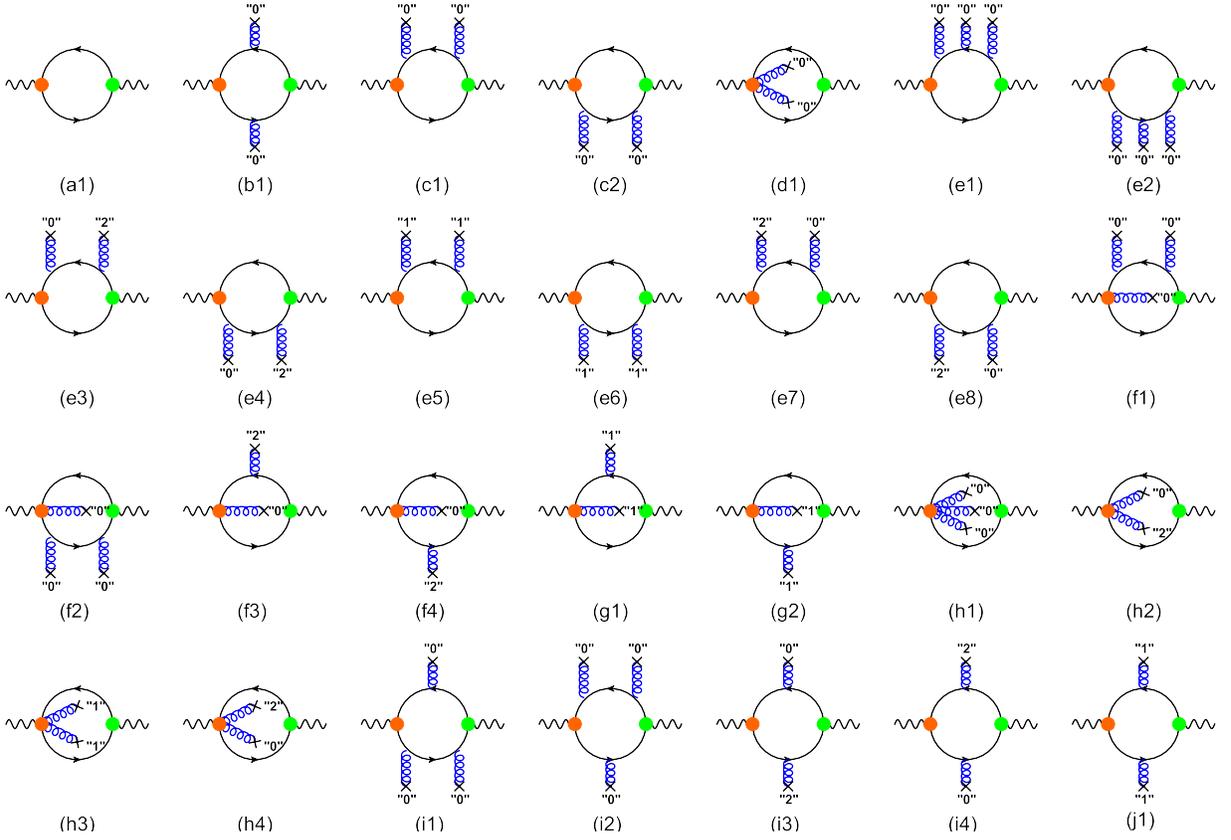}
\caption{Feynman diagrams for the moments of the HP leading-twist DA. The left big dot and the right big dot stand for the vertex operators $\not\! z \gamma_5 (z\cdot \tensor{D})^n$ and $\not\! z \gamma_5$ in the currents $J_n(x)$ and $J^\dagger_0(0)$, respectively. The cross symbol attached to the gluon line indicates the tensor of the local gluon background field, and ``$n$" indicates $n_{\rm th}$-order covariant derivative.}
\label{feynDA}
\end{figure*}

Fig.\ref{feynDA} shows the corresponding Feynman diagrams for deriving the moments $\left<\xi^n\right>_{\rm HP}$. In Fig.\ref{feynDA}, the left big dot and the right big dot stand for the vertex operators $\not\! z \gamma_5 (z\cdot \tensor{D})^n$ and $\not\! z \gamma_5$ in the currents $J_n(x)$ and $J^\dagger_0(0)$, respectively; the cross symbol attached to the gluon line indicates the tensor of the local gluon background field, and ``$n$" indicates $n_{\rm th}$-order covariant derivative. In different to Fig.\ref{feynDC}, there are seven Feynman diagrams that have not been shown in Fig.\ref{feynDA}, because they have no contribution for the moments $\left<\xi^n\right>_{\rm HP}$ due to their quark loops explicitly lead to $\textrm{Tr}[\cdots]=0$. Fig.\ref{feynDA}.(a1) provides the perturbative contribution, Fig.\ref{feynDA}.(b1-d1) provide the double-gluon condensate contribution and the remaining twenty-three diagrams provide the triple-gluon condensate contribution. Furthermore, comparing with Fig.\ref{feynDC}, we have some extra diagrams for the present case, i.e. Fig.\ref{feynDA}.(d1), Figs.\ref{feynDA}.(f1-h4), which are due to the new vertex operator $(z\cdot \tensor{D})^n$.

Following the standard SVZ procedures of the sum rules, the final sum rules for the moments of the HP leading-twist DA can be written as
\begin{widetext}
\begin{eqnarray}
\frac{f_{\rm HP}^2 \left<\xi^n\right>_{\rm HP}}{M^2 \exp \left[ m_{\rm HP}^2/M^2 \right]} = \frac{1}{\pi} \frac{1}{M^2} \int^{s_{\rm HP}}_{t_{\rm min}} ds e^{-s/M^2} \textrm{Im} I_{\rm pert} (s) + \hat{L}_M I_{\left<G^2\right>}(Q^2) + \hat{L}_M I_{\left<G^3\right>}(Q^2),
\label{srborel}
\end{eqnarray}
where
\begin{eqnarray}
\textrm{Im} I_{\rm pert}(s) &=& \frac{3}{8\pi(n+1)(n+3)} \left\{ \left( \frac{s_1+\bar{s}v}{s} - 1 \right)^{n+1} \left[ (n+1) \frac{(s_1+\bar{s}v) (-s_2+\bar{s}v)}{2s^2} - 1 \right] - ( v \to -v ) \right\}, \\
\hat{L}_M I_{\left<G^2\right>}(Q^2) &=& \frac{\left<\alpha_s G^2\right>}{6\pi} \int^1_0 dx \exp \left[ -\frac{m_1^2x + m_2^2(1-x)}{M^2x(1-x)} \right] \left\{ \left[ \frac{1}{2} (2x-1)^n x^2(1-x)^2 + n(n-1) (2x-1)^{n-2} \right.\right. \nonumber\\
&\times& \left.\left. x^3(1-x)^3 \right] \frac{1}{M^4x^2(1-x)^2} - (2x-1)^n x(1-x) \frac{m_1^2x^3 + m_2^2(1-x)^3}{2M^6x^3(1-x)^3} \right\}, \\
\hat{L}_M I_{\left<G^3\right>}(Q^2) &=& \frac{\left<g_s^3 f G^3\right>}{(4\pi)^2} \int^1_0 dx \exp \left[ -\frac{m_1^2x + m_2^2(1-x)}{M^2x(1-x)} \right] \left\{ \left[ -\frac{45}{8} (2x-1)^n x(1-x) (x^3+(1-x)^3) \right.\right. \nonumber\\
&-& (2x-1)^n x^2(1-x)^2 \left( \frac{16n}{9}x(1-x) + \frac{22n+69}{72} \right) - \frac{n(n-1)}{9} (2x-1)^{n-2} x^3(1-x)^3 \nonumber\\
&\times& \left. ((n+1)x(1-x) + 16x^2+16(1-x)^2) \right] \frac{1}{2M^6x^3(1-x)^3} + \left[ -\frac{3}{4} (2x-1)^n x(1-x) \right. \nonumber\\
&\times& (m_1^2x^4 + m_2^2(1-x)^4) + \frac{11n}{6} (2x-1)^{n-1} x^2(1-x)^2 (m_1^2x^3 - m_2^2(1-x)^3) - \frac{n(n-1)}{3} (2x-1)^{n-2} \nonumber\\
&\times& x^4(1-x)^4 (m_1^2x + m_2^2(1-x)) + (2x-1)^n x^2(1-x)^2 \left( -\frac{23}{12} (m_1^2x^2 + m_2^2(1-x)^2)
\right. \nonumber\\
&+& \left. \frac{1}{3} (m_1m_2 - 6m_1^2 - 6m_2^2) x(1-x) + 2(m_1^2x + m_2^2(1-x)) \right) - \frac{8n}{3} (2x-1)^n x^3(1-x)^3 \nonumber\\
&\times& \left.\left. (m_1^2x + m_2^2(1-x)) \right] \frac{1}{6M^8x^4(1-x)^4} + \frac{2}{5} (2x-1)^n x(1-x) \frac{m_1^4x^5 + m_2^4(1-x)^5}{24M^{10}x^5(1-x)^5} \right\}.
\end{eqnarray}
\end{widetext}

Up to $6_{\rm th}$-order, the moments $\left<\xi^n\right>_{\rm HP}$ and the Gegenbauer moments $a_n^{\rm HP}$ at the same scale $\mu$ can be related via the following equations:
\begin{eqnarray}
\left<\xi^1\right>_{\rm HP}|_\mu &=& \frac{3}{5} a_1^{\rm HP}(\mu), \label{rel_mom1}\\
\left<\xi^2\right>_{\rm HP}|_\mu &=& \frac{1}{5} + \frac{12}{35} a_2^{\rm HP}(\mu), \label{rel_mom2}\\
\left<\xi^3\right>_{\rm HP}|_\mu &=& \frac{9}{35} a_1^{\rm HP}(\mu) + \frac{4}{21} a_3^{\rm HP}(\mu), \label{rel_mom3} \\
\left<\xi^4\right>_{\rm HP}|_\mu &=& \frac{3}{35} + \frac{8}{35} a_2^{\rm HP}(\mu) + \frac{8}{77} a_4^{\rm HP}(\mu), \label{rel_mom4} \\
\left<\xi^5\right>_{\rm HP}|_\mu &=& \frac{1}{7} a_1^{\rm HP}(\mu) + \frac{40}{231} a_3^{\rm HP}(\mu) + \frac{8}{143} a_5^{\rm HP}(\mu), \label{rel_mom5} \\
\left<\xi^6\right>_{\rm HP}|_\mu &=& \frac{1}{21} + \frac{12}{77} a_2^{\rm HP}(\mu) + \frac{120}{1001} a_4^{\rm HP}(\mu) \nonumber \\
&& +\frac{64}{2145} a_6^{\rm HP}(\mu).   \label{rel_mom6}
\end{eqnarray}
Thus, inversely, we can derive the Gegenbauer moments $a_n^{\rm HP}$ from the above sum rules for $\left<\xi^n\right>_{\rm HP}$. Usually, the Gegenbauer moments $a_n^{\rm HP}$ are known for an initial scale $\mu_0$ around $\Lambda_{\rm QCD}$, which can be evolved from any scale $\mu$ via the equation
\begin{eqnarray}
a^{\rm HP}_n(\mu) = \left( \frac{\alpha_s(\mu)}{\alpha_s(\mu_0)} \right)^{\frac{\epsilon_n}{4\pi b_0}} a^{\rm HP}_n(\mu_0),
\label{SEE}
\end{eqnarray}
where
\begin{displaymath}
\epsilon_n =\frac{4}{3} \left( 1 - \frac{2}{(n+1)(n+2)} + 4 \sum^{n+1}_{j=2} \frac{1}{j} \right).
\end{displaymath}
For the running coupling, we adopt~\cite{PDG}
\begin{eqnarray}
\alpha_s(\mu) &=& \frac{1}{b_0 t} \left[ 1 - \frac{b_1}{b_0^2} \frac{\ln t}{t}
\right. \nonumber\\
&& \quad\quad\quad\left. +\frac{b_1^2 (\ln^2 t - \ln t - 1) + b_0 b_2}{b_0^4 t^2} \right]
\label{Coupling}
\end{eqnarray}
with $t= \ln \frac{\mu^2}{\Lambda^2_{\rm QCD}}$,
\begin{eqnarray}
b_0 &=& \frac{33-2n_f}{12\pi},
\nonumber\\
b_1 &=& \frac{153 - 19n_f}{24\pi^2},
\nonumber\\
b_2 &=& \frac{2857 - \frac{5033}{9}n_f + \frac{325}{27}n_f^2}{128\pi^2}.
\nonumber
\end{eqnarray}

\subsection{A Model for the HP Leading-Twist DAs}

The meson DA can be derived from its light-cone wavefunction by integrating out its transverse components. Thus, it is helpful to construct a HP leading-twist wavefunction and then get its DA. For the purpose, one may assume that the HPs wavefunctions have similar form as those of the pseudoscalars kaon with $SU_{f}(3)$-breaking effect~\cite{KDA} and the $D$ meson or $B$ meson~\cite{TFF,BDDA}. Based on the BHL-prescription~\cite{BHL}, the HP wavefunction can be constructed as
\begin{eqnarray}
\Psi_{\rm HP}(x,\textbf{k}_\bot) = \chi_{\rm HP}(x,\textbf{k}_\bot) \Psi_{\rm HP}^R (x,\textbf{k}_\bot),  \label{etacWF}
\end{eqnarray}
where $\textbf{k}_\bot$ is the transverse momentum, $\chi_{\rm HP}(x,\textbf{k}_\bot)$ is the spin-space wavefunction and $\Psi_{\rm HP}^R (x,\textbf{k}_\bot)$ stands for the spatial wavefunction. The spin-space wavefunction $\chi_{\rm HP}(x,\textbf{k}_\bot)$ takes the form~\cite{PETAC}
\begin{eqnarray}
\chi_{\rm HP}(x,\textbf{k}_\bot) = \frac{\hat{m}_1(1-x) + \hat{m}_2x}{\sqrt{\textbf{k}_\bot^2 + [\hat{m}_1(1-x) + \hat{m}_2x]^2}},
\label{sswf}
\end{eqnarray}
where $\hat{m}_{1,2}$ are the constituent quark masses for the HP. $\hat{m}_{1} = \hat{m}_{b}$ and $\hat{m}_{2} = \hat{m}_{c}$ for the case of $B_c$ meson, $\hat{m}_{1}=\hat{m}_{2} = \hat{m}_{c}$ ($\hat{m}_{b}$) for the case of $\eta_{c}$ ($\eta_{b}$). We take $\hat{m}_{c} = 1.8 \textrm{GeV}$ and $\hat{m}_b = 4.7 \textrm{GeV}$ to do our numerical calculations. It is noted that different choices of $\hat{m}_{c}$ or $\hat{m}_{b}$ will lead to quite small differences to the HP DA. Because $\hat{m}_b, \hat{m}_c >> \Lambda_{QCD}$, the spin-space wavefunction $\chi_{\rm HP}$ tends to $1$ for the heavy scalars, thus, one may omit such factor as a simplified model. The spatial wavefunction $\Psi_{\rm HP}^R (x,\textbf{k}_\bot)$ takes the form
\begin{eqnarray}
\Psi_{\rm HP}^R (x,\textbf{k}_\bot) &=& A_{\rm HP} \varphi_{\rm HP}(x) \times \nonumber\\
&& \!\!\!\! \exp \left[ \frac{-1}{8\beta_{\rm HP}^2} \left( \frac{\textbf{k}_\bot^2 + \hat{m}_1^2}{x} + \frac{\textbf{k}_\bot^2 + \hat{m}_2^2}{1-x} \right) \right],   \label{etacRWF}
\end{eqnarray}
where $A_{\rm HP}$ is normalization constant. The parameter $\beta_{\rm HP}$ is a harmonious parameter that dominantly determines the wavefunction transverse distributions. The function $\varphi_{\rm HP}(x)$ dominantly dominates the wavefunction's longitudinal distribution, whose behavior is further dominated by its first several Gegenbauer polynomials. By keeping up to $6_{\rm th}$-order Gegenbauer moments, it can be expansion as
\begin{eqnarray}
\varphi_{\rm HP}(x) = 1 + \sum^6_{n=1} B_n^{\rm HP} \times C^{3/2}_n(2x-1),
\label{varphi}
\end{eqnarray}
in which $B_{1,3,5}^{\rm HP}$ should be $0$ for the case of $\eta_c$ or $\eta_b$ DA, due to the fact that the $\eta_c$ or $\eta_b$ DA should be unchanged over the transformation $x \leftrightarrow (1-x)$.

Using the relationship between the HP leading-twist DA and the HP wavefunction,
\begin{eqnarray}
\phi_{\rm HP}(x,\mu) = \frac{2\sqrt{6}}{f_{\rm HP}} \int_{|\textbf{k}_\bot|^2 \leq \mu_0^2} \frac{d^2\textbf{k}_\bot}{16 \pi^3} \Psi_{\rm HP}(x,\textbf{k}_\bot),    \label{rel_dawf}
\end{eqnarray}
we can obtain the required leading-twist DA for the HP. That is, after integrating over the transverse momentum for the wavefunction (\ref{etacWF}), we obtain
\begin{widetext}
\begin{eqnarray}
\phi_{\rm HP}(x,\mu_0) &=& \frac{\sqrt{3} A_{\rm HP} \tilde{m} \beta_{\rm HP}} {2\pi^{3/2} f_{\rm HP}} \sqrt{x(1-x)} \varphi_{\rm HP}(x) \times \exp \left[ -\frac{\hat{m}_1^2(1-x) + \hat{m}_2^2x - \tilde{m}^2}{8\beta_{\rm HP}^2 x(1-x)} \right] \nonumber\\
&& \times\left\{ \textrm{Erf} \left[ \sqrt{\frac{\tilde{m}^2 + \mu_0^2}{8\beta_{\rm HP}^2 x(1-x)}} \right] - \textrm{Erf} \left[ \sqrt{\frac{\tilde{m}^2}{8\beta_{\rm HP}^2 x(1-x)}} \right] \right\},
\label{etacDA}
\end{eqnarray}
\end{widetext}
where $\tilde{m} = \hat{m}_1(1-x) + \hat{m}_2x$, $\mu_0\sim\Lambda_{QCD}$ is the factorization scale, and the error function $\textrm{Erf}(x) = \frac{2}{\sqrt{\pi}} \int^x_0 e^{-t^2} dt$.

The wavefunction parameters $A_{\rm HP}$, $B_n^{\rm HP}$ and $\beta_{\rm HP}$ can be determined by the following constraints:
\begin{itemize}
\item The normalization condition,
\begin{eqnarray}
\int^1_0 dx \int_{|\textbf{k}_\bot|^2 \leq \mu_0^2} \frac{d^2 \textbf{k}_\bot}{16\pi^3} \Psi_{\rm HP}(x,\textbf{k}_\bot) = \frac{f_{\rm HP}}{2\sqrt{6}}. \label{NC}
\end{eqnarray}
The decay constant $f_{\rm HP}$ can be determined by the sum rules (\ref{f_srborel}).

\item The probability of finding the leading Fock state $\left.\left.\right|Q_1\bar{Q}_2\right>$ in the HP Fock state expansion,
\begin{eqnarray}
P_{\rm HP} = \int^1_0 dx \int \frac{d^2 \textbf{k}_\bot}{16\pi^3} \left| \Psi^R_{\rm HP} (x,\textbf{k}_\bot) \right|^2.   \label{P}
\end{eqnarray}
Equivalently, one can replace the constraint (\ref{P}) by the average value of the squared transverse momentum $\left<\textbf{k}_\bot^2\right>_{\rm HP}$, which is measurable and is defined as
\begin{displaymath}
\left<\textbf{k}_\bot^2\right>_{\rm HP} = \int^1_0 dx \int \frac{d^2 \textbf{k}_\bot}{16\pi^3} |\textbf{k}_\bot|^2 \frac{ \left|\Psi^R_{\rm HP} (x,\textbf{k}_\bot) \right|^2}{P_{\rm HP}}.
\end{displaymath}
The experimental measurements on $\left<\textbf{k}_\bot^2\right>_{\rm HP}$ are not available at the present. We adopt the constraint (\ref{P}) and take $P_{\eta_c} \simeq 0.8$~\cite{ECDA_BHL,PETAC} and $P_{B_c}\sim P_{\eta_b} \simeq 1$~\cite{BDDA} to do the calculation. The choice of $P_{\eta_b}\sim P_{B_c} > P_{\eta_c}$ is reasonable, since with the increase of the constituent quark masses, the valence Fock state occupies a bigger fraction in hadron and the probability of finding the valence Fock state will be close to unity in the non-relativistic limit. We have checked that all the wavefunction parameters change very slightly by varying $P_{B_c}$ from $1.0$ to $0.9$, which indicates that the $B_c$ meson already reaches the non-relativistic limit.

\item The Gegenbauer moments can also be derived from the following definition
\begin{eqnarray}
a_n^{\rm HP}(\mu_0) = \frac{\int^1_0 dx \phi_{\rm HP}(x,\mu_0) C^{3/2}_n(2x-1)}{\int^1_0 dx 6x(1-x) \left[ C^{3/2}_n(2x-1) \right]^2}.
\label{GMD}
\end{eqnarray}
They should be equal to the Gegenbauer moments determined from the values of $\left<\xi^n\right>_{\rm HP}$, which can be determined from the sum rules (\ref{srborel}).

\end{itemize}

Using these constraints, one can strictly determine the wavefunction parameters $A_{\rm HP}$, $B_n^{\rm HP}$ and $\beta_{\rm HP}$ at an initial scale $\mu_0$. These parameters are scale dependent, one can obtain their values at any scale $\mu$ via the following evolution equation~\cite{EE}
\begin{widetext}
\begin{eqnarray}
x_1 x_2 \mu^2 \frac{\partial \tilde{\phi}_{\rm HP}(x_i,\mu)}{\partial \mu^2} = C_F \frac{\alpha_s(\mu^2)}{4\pi} \left\{ \int^1_0 [dy] V(x_i,y_i) \tilde{\phi}_{\rm HP}(y_i,\mu) - x_1 x_2 \tilde{\phi}_{\rm HP}(x_i,\mu) \right\},
\label{ee}
\end{eqnarray}
\end{widetext}
where $C_F = 4/3$,
\begin{eqnarray}
[dy] &=& dy_1 dy_2 \delta(1-y_1-y_2), \nonumber\\
V(x_i,y_i) &=& 2 \Big[ x_1y_2 \theta(y_1-x_1)\times \nonumber\\
&& \left( \delta_{h_1\bar{h}_2} + \frac{\Delta}{(y_1-x_1)} \right) + (1 \leftrightarrow 2) \Big], \nonumber\\
\phi_{\rm HP}(x_i,\mu) &=& x_1x_2 \tilde{\phi}_{\rm HP}(x_i,\mu), \nonumber\\
\Delta \tilde{\phi}_{\rm HP}(y_i,\mu) &=& \tilde{\phi}_{\rm HP}(y_i,\mu) - \tilde{\phi}_{\rm HP}(x_i,\mu),
\nonumber
\end{eqnarray}
$\delta_{h_1\bar{h}_2} = 1$ when the $Q_1$ and $\bar{Q}_2$ have opposite helicities and $\delta_{h_1\bar{h}_2} = 0$ for other cases.

\section{numerical analysis}

\subsection{Input parameters}

To determine the HP decay constants and the first several moments of the HP leading-twist DA, we take~\cite{PDG}
\begin{eqnarray}
m_{\eta_c} &=& (2.9837 \pm 0.0007) \textrm{GeV},
\nonumber\\
m_{B_c} &=& (6.2745 \pm 0.0018) \textrm{GeV},
\nonumber\\
m_{\eta_b} &=& (9.3980 \pm 0.0032) \textrm{GeV},
\nonumber\\
\bar{m}_c (\bar{m}_c) &=& (1.275 \pm 0.025) \textrm{GeV},
\nonumber\\
\bar{m}_b (\bar{m}_b) &=& (4.18 \pm 0.03) \textrm{GeV}.
\label{inputs}
\end{eqnarray}
The $\overline{MS}$ $c$- and $b$-quark masses at any other scale can be derived from the evolution~\cite{PDG}
\begin{eqnarray}
\bar{m}_c(\mu) &=& \bar{m}_c(\bar{m}_c) \left[ \frac{\alpha_s(\mu)}{\alpha_s(\bar{m}_c)} \right]^{\frac{12}{25}},
\nonumber\\
\bar{m}_b(\mu) &=& \bar{m}_b(\bar{m}_b) \left[ \frac{\alpha_s(\mu)}{\alpha_s(\bar{m}_b)} \right]^{\frac{12}{23}},
\label{MassRGE}
\end{eqnarray}
From $\alpha_s(m_Z) = 0.1184 \pm 0.0007$ with $m_Z = (91.1876 \pm 0.0021) \textrm{GeV}$~\cite{PDG}, we predict $\Lambda_{\rm QCD} \simeq 270\textrm{MeV}$, $257 \textrm{MeV}$ and $204 \textrm{MeV}$ for the flavor $n_f = 3$, $4$ and $5$, respectively. We take the scale-independence dimension-four gluon condensate $\left<\alpha_s G^2\right> = (0.038 \pm 0.011) \textrm{GeV}^4$~\cite{GLUCON} and $\left<g_s^3 f G^3\right> = (0.013 \pm 0.007) \textrm{GeV}^6$~\cite{BFSR}.

\subsection{The HP Decay Constants}

\begin{table}[htb]
\caption{The HP decay constants for $s_{\eta_c}=18\textrm{GeV}^2$, $s_{B_c}=45\textrm{GeV}^2$ and $s_{\eta_b}=90\textrm{GeV}^2$ under the allowable Borel windows, where all the other input parameters are taken to be their central values.}
\begin{tabular}{ c | c c c }
\hline
~HP~&~$\eta_c$~& ~$B_c$~ & ~$\eta_b$~ \\
\hline
~$s_{\rm HP}(\textrm{GeV}^2)$~& ~$18$~ & ~$45$~ & ~$90$~ \\
~$M^2(\textrm{GeV}^2)$~& ~$[2, 11]$~ & ~$[9,13]$~ & ~$[16,20]$~ \\
~$f_{\rm HP}(\textrm{MeV})$~& ~$453 \pm 3$~ & ~$498 \pm 9$~ & ~$811 \pm 9$~ \\
\hline
\end{tabular}
\label{tfhps}
\end{table}

\begin{figure}[htb]
\centering
\includegraphics[width=0.45\textwidth]{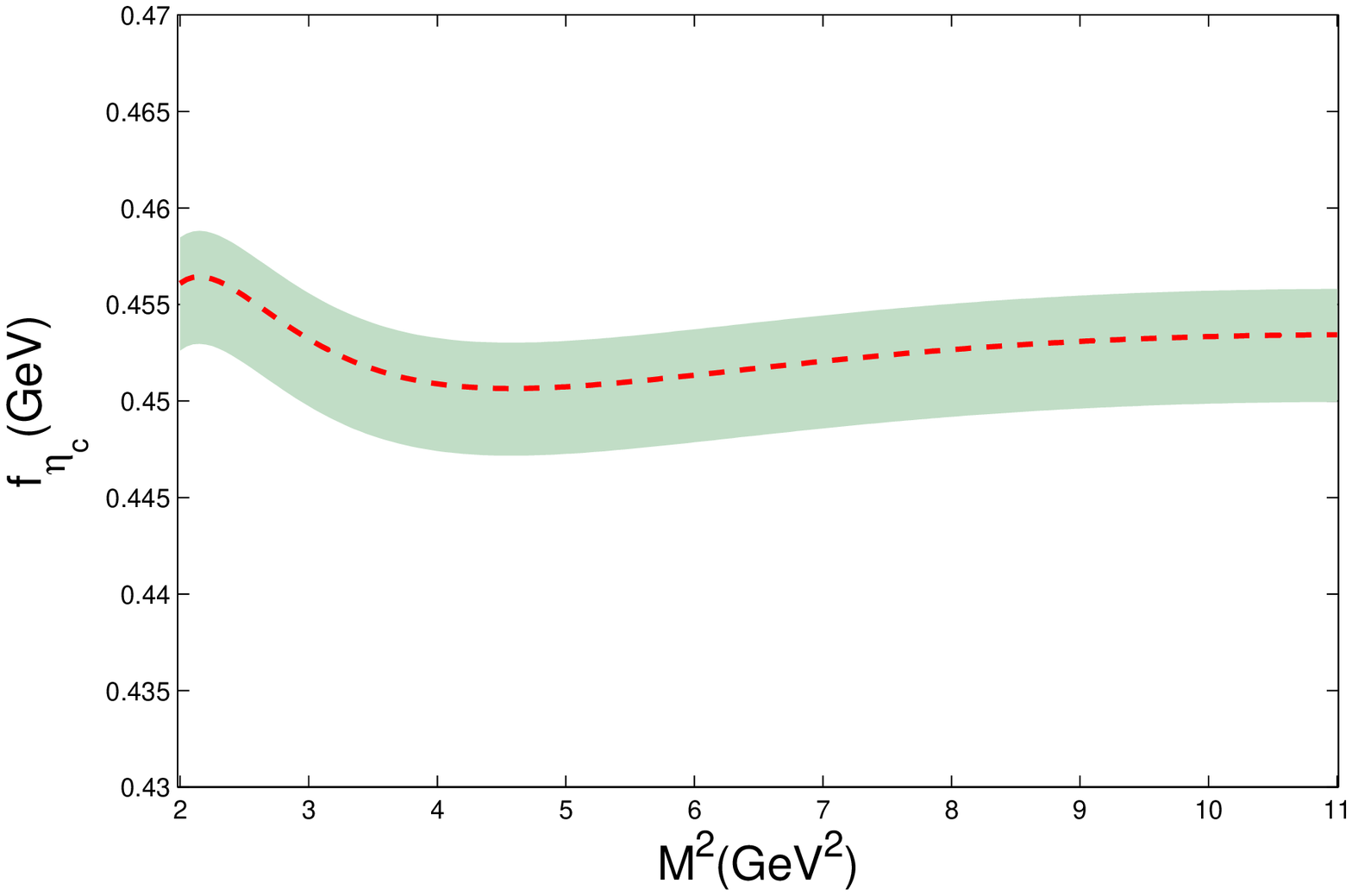}
\includegraphics[width=0.45\textwidth]{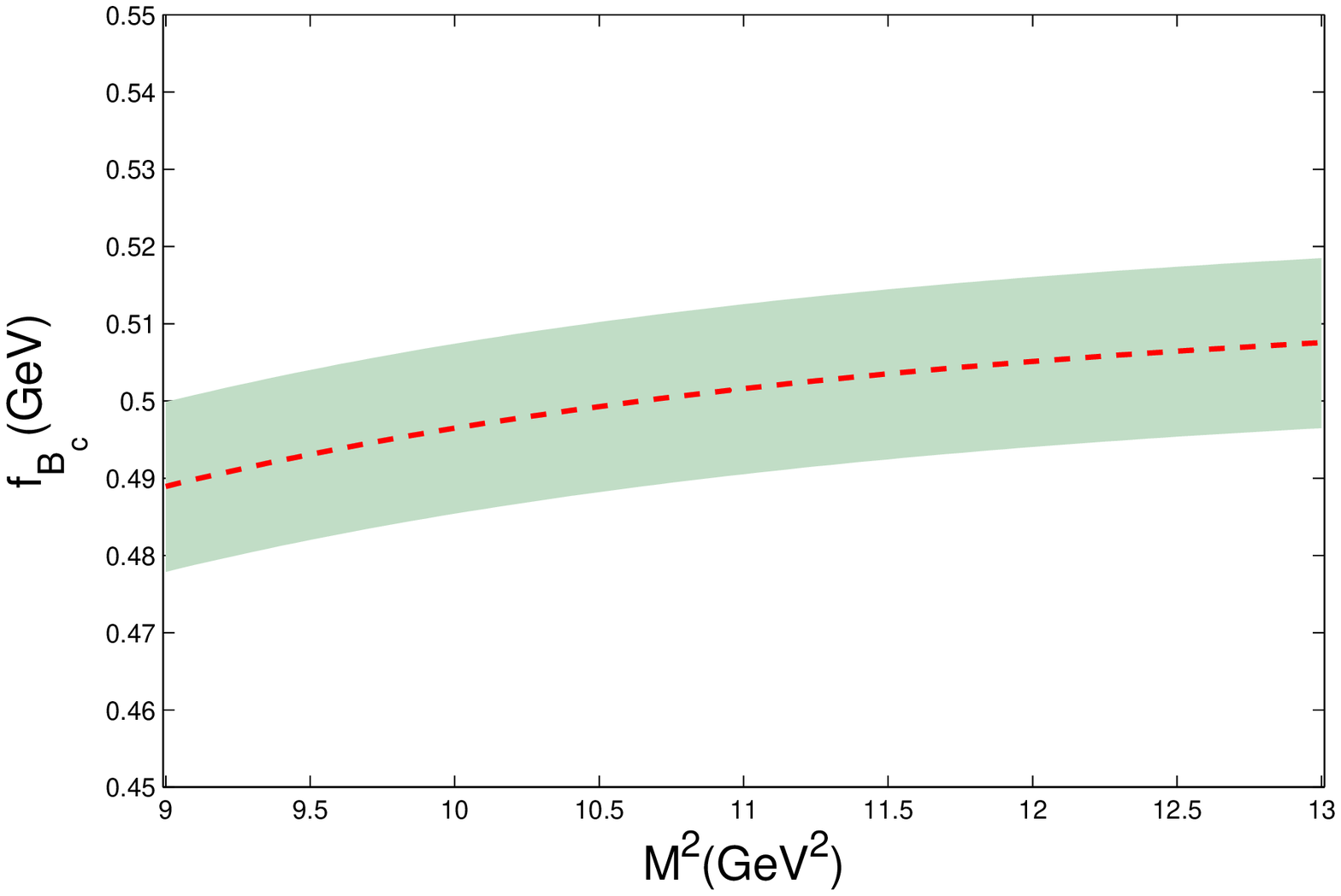}
\includegraphics[width=0.45\textwidth]{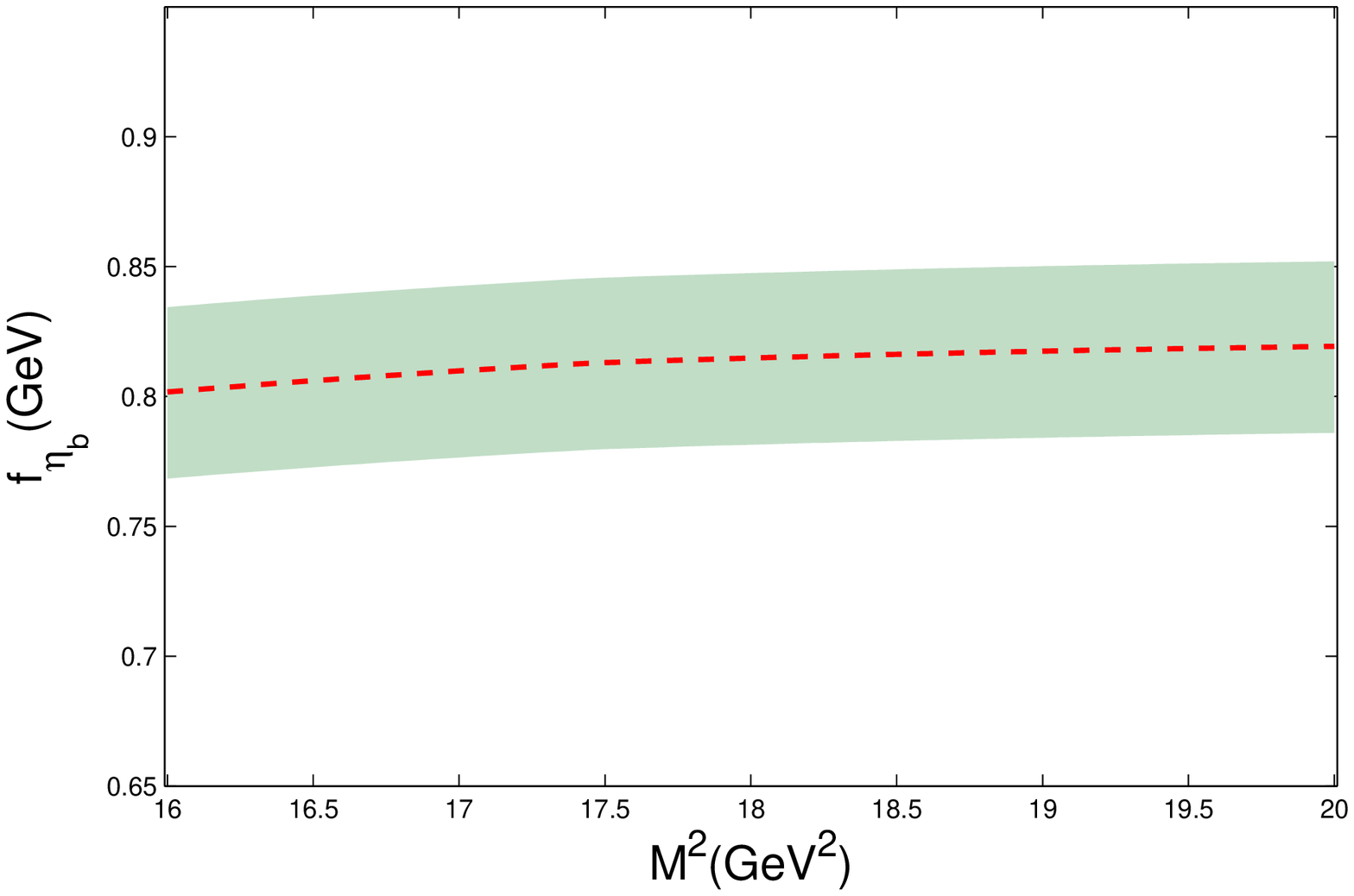}
\caption{The HP decay constants versus the Borel parameter $M^2$. The shaded band indicates the uncertainty.}
\label{ffhp}
\end{figure}

\begin{table}[htb]
\caption{A comparison of our present HP decay constants (in unit $\textrm{MeV}$) with those obtained under various approaches~\cite{FHP85, FHP88, FHP89, FHP91, FHP92, FHP931, FHP932, FHP941, FHP942, FHP96, FHP97, FHP01, FHP04, FHP061, FHP062, FHPL}. }
\begin{tabular}{ c c c | c }
\hline
~$f_{\eta_c}$~& ~$f_{B_c}$~ & ~$f_{\eta_b}$~ & ~$\textrm{Ref.}$~ \\
\hline
~$453 \pm 4$~& ~$498 \pm 14$~ & ~$811 \pm 34$~ & ~$\textrm{This work}$~ \\
~$/$~& ~$510 \pm 50$~ & ~$/$~ & ~\cite{FHP85}~ \\
~$/$~& ~$400 \pm 20$~ & ~$/$~ & ~\cite{FHP88}~ \\
~$/$~& ~$375 \pm 40$~ & ~$/$~ & ~\cite{FHP89}~ \\
~$/$~& ~$570 \pm 60$~ & ~$/$~ & ~\cite{FHP91}~ \\
~$/$~& ~$300$~ & ~$/$~ & ~\cite{FHP92}~ \\
~$320 \pm 40$~& ~$360 \pm 60$~ & ~$500 \pm 100$~ & ~\cite{FHP931}~ \\
~$/$~& ~$570 \pm 60$~ & ~$/$~ & ~\cite{FHP932}~ \\
~$/$~& ~$383 \pm 27$~ & ~$/$~ & ~\cite{FHP941}~ \\
~$/$~& ~$300 \pm 65$~ & ~$/$~ & ~\cite{FHP942}~ \\
~$/$~& ~$385 \pm 25$~ & ~$/$~ & ~\cite{FHP96}~ \\
~$420 \pm 52$~& ~$/$~ & ~$705 \pm 27$~ & ~\cite{FHP97}~ \\
~$/$~& ~$400 \pm 45$~ & ~$/$~ & ~\cite{FHP01}~ \\
~$/$~& ~$395 \pm 15$~ & ~$/$~ & ~\cite{FHP04}~ \\
~$484$~& ~$399$~ & ~$/$~ & ~\cite{FHP061}~ \\
~$490$~& ~$/$~ & ~$/$~ & ~\cite{FHP062}~ \\
~$438 \pm 8$~& ~$489 \pm 5$~ & ~$801 \pm 9$~ & ~\cite{FHPL}~ \\
\hline
\end{tabular}
\label{tfhp}
\end{table}

To set the threshold parameter $s_{\rm HP}$ and the allowable Borel window for the sum rules (\ref{f_srborel}), we require that the continuum contribution to be less than $30\%$, and the values for $f_{\rm HP}$ are stable in the Borel window. We obtain $s_{\eta_c}=18\textrm{GeV}^2$, $s_{B_c}=45\textrm{GeV}^2$ and $s_{\eta_b}=90\textrm{GeV}^2$. Our predictions for the HP decay constants $f_{\rm HP}$ under the allowable Borel windows are put in Table \ref{tfhps}, where all other input parameters are taken as their central values. We put the curves for the decay constants $f_{\eta_c}$, $f_{B_c}$ and $f_{\eta_b}$ versus the Borel parameter $M^2$ in Fig.\ref{ffhp}, where the shaded bands indicate the uncertainties from the input parameters $m_{\rm HP}$, $m_{c,b}$, $\left<\alpha_s G^2\right>$ and $\left<g_s^3 fG^3\right>$. By taking all uncertainty errors into consideration and adding them in quadrature, our final predictions on $f_{\rm HP}$ are put in Table \ref{tfhp}. As a comparison, some typical estimations on the HP decay constants derived under various approaches~\cite{FHP85, FHP88, FHP89, FHP91, FHP92, FHP931, FHP932, FHP941, FHP942, FHP96, FHP97, FHP01, FHP04, FHP061, FHP062, FHPL} are also presented. Table \ref{tfhp} shows that our present estimations on HP decay constants agree with those derived under the Lattice QCD~\cite{FHPL}, especially for $f_{B_c}$ and $f_{\eta_b}$.

\subsection{The HP Leading-Twist DAs}

\begin{table}[htb]
\caption{The HP leading-twist DA moments $\left<\xi^n\right>_{\rm HP}$ up to $6_{\rm th}$-order. The errors are squared average of those from all the input parameters, such as the Borel parameter, the condensates and the bound state parameters. The scale $\mu$ is set to be $\bar{m}_c(\bar{m}_c)$ for $\eta_c$ and $\bar{m}_b(\bar{m}_b)$ for $B_c$ and $\eta_b$. }
\begin{tabular}{ c | c c c }
\hline
~ & ~$\eta_c(\mu = \bar{m}_c(\bar{m}_c))$~ & ~$B_c(\mu = \bar{m}_b(\bar{m}_b))$~ & ~$\eta_b(\mu = \bar{m}_b(\bar{m}_b))$~ \\
\hline
~$\left<\xi^1\right>$~& ~$     0     $~ & ~$0.279 \pm 0.023$~ & ~$0$~ \\
~$\left<\xi^2\right>$~& ~$0.073 \pm 0.009$~ & ~$0.182 \pm 0.005$~ & ~$0.067 \pm 0.007$~ \\
~$\left<\xi^3\right>$~& ~$     0     $~ & ~$0.100 \pm 0.006$~ & ~$0$~ \\
~$\left<\xi^4\right>$~& ~$0.014 \pm 0.003$~ & ~$0.071 \pm 0.003$~ & ~$0.011 \pm 0.002$~ \\
~$\left<\xi^5\right>$~& ~$    0      $~ & ~$0.047 \pm 0.002$~ & ~$0$~ \\
~$\left<\xi^6\right>$~& ~$0.004 \pm 0.001$~ & ~$0.036 \pm 0.001$~ & ~$0.003 \pm 0.001$~ \\
\hline
\end{tabular}
\label{thpxiBW}
\end{table}

\begin{figure}[htb]
\centering
\includegraphics[width=0.45\textwidth]{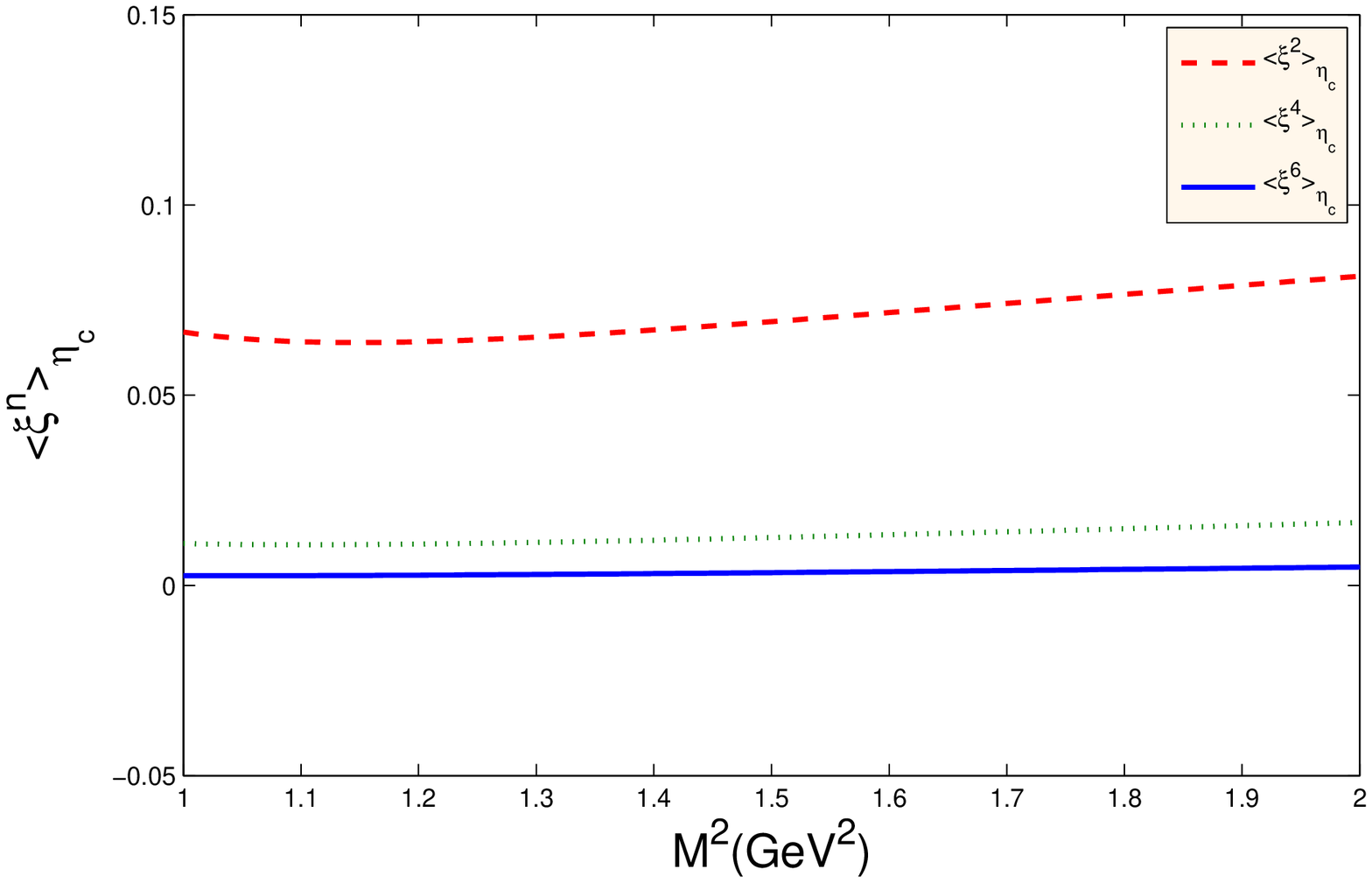}
\includegraphics[width=0.45\textwidth]{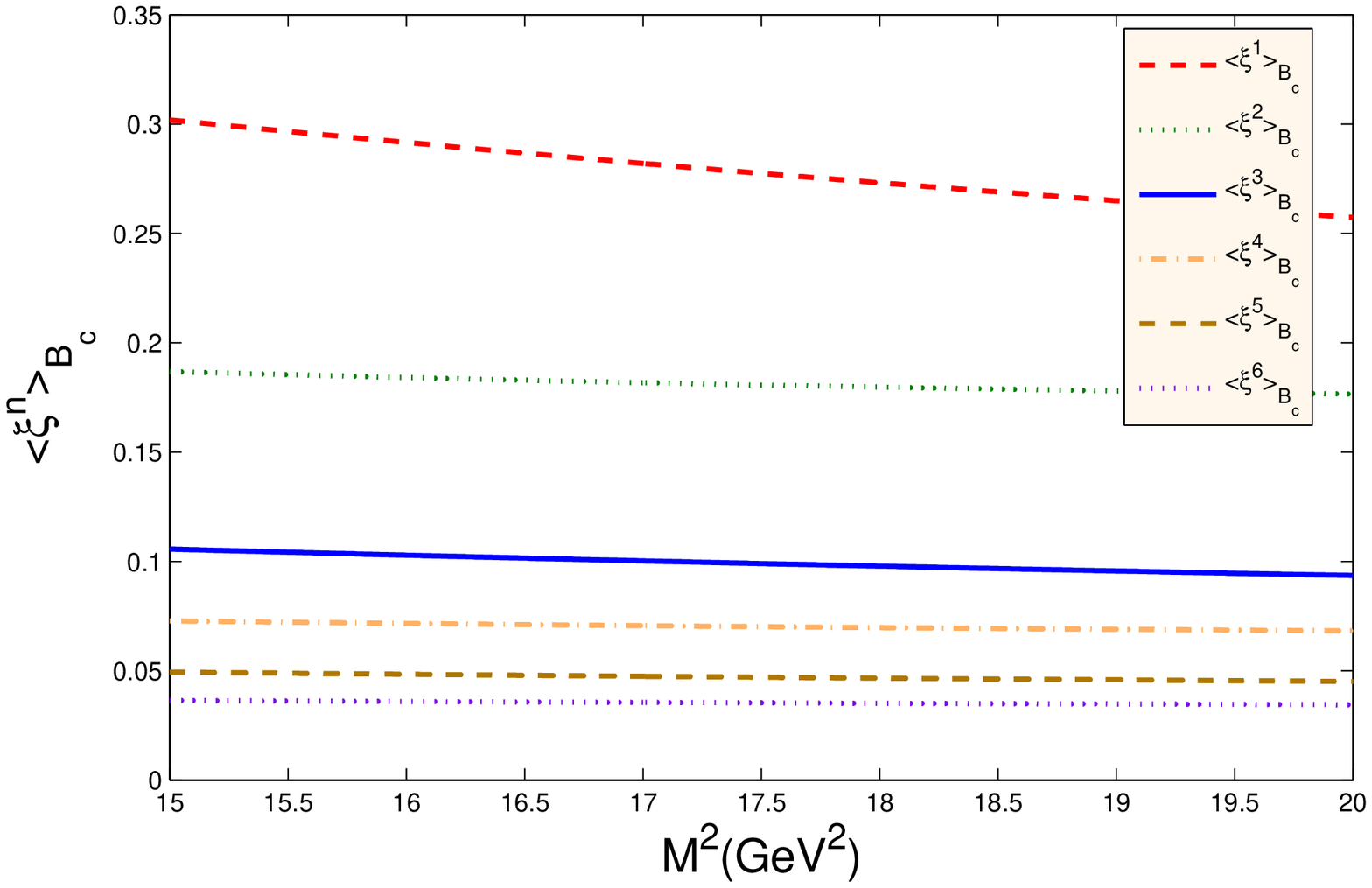}
\includegraphics[width=0.45\textwidth]{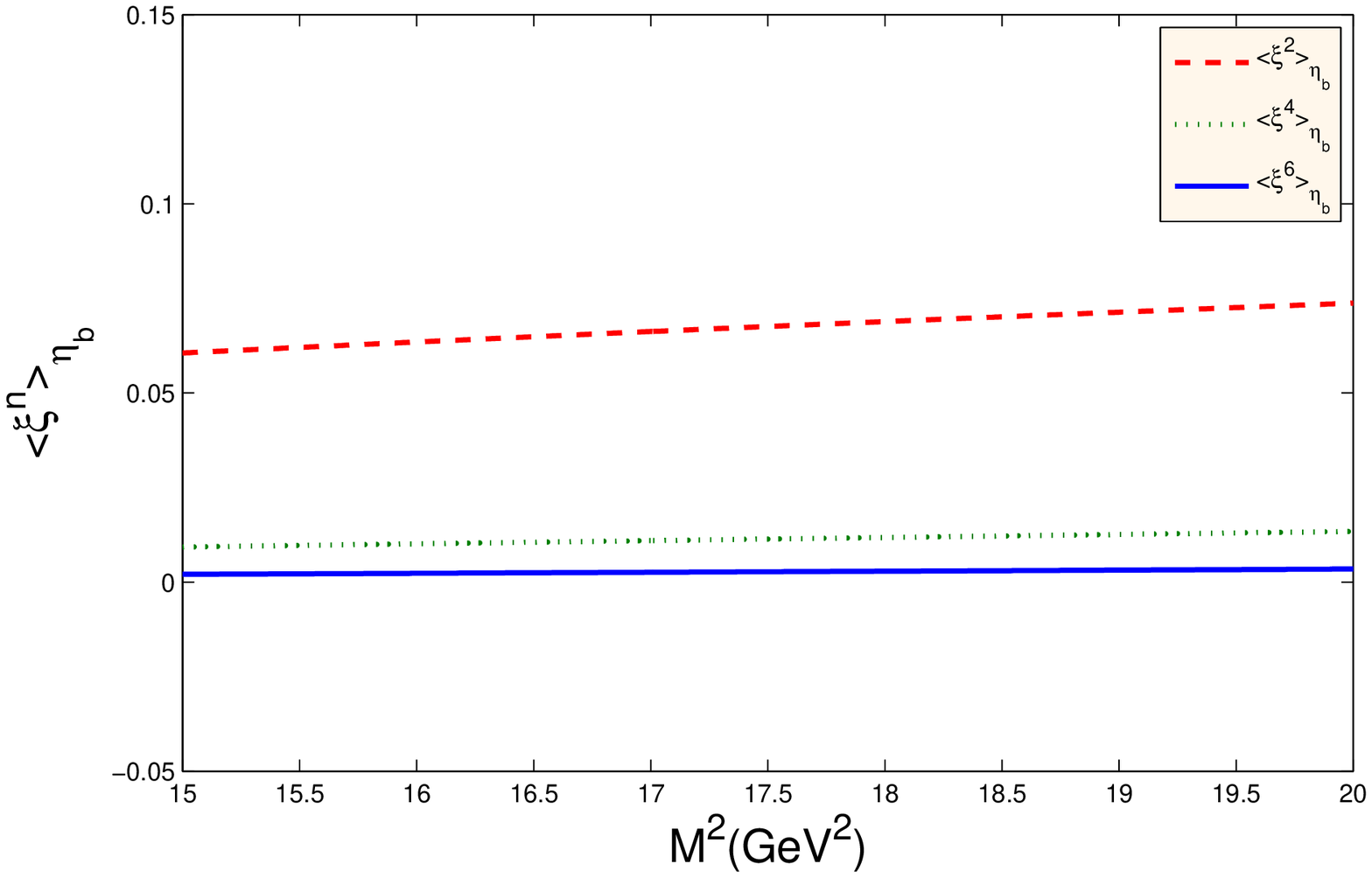}
\caption{The first several moments $\left<\xi^n\right>_{\rm HP}$ versus the Borel parameter $M^2$. Where the input parameters are taken as the central values.}
\label{fhpxin}
\end{figure}

First, we calculate the HP leading-twist DA moments $\left<\xi^n\right>_{\rm HP}$ with the SVZ sum rules (\ref{srborel}). As suggested by Braguta etal.~\cite{ECDA_BLL}, we set the continue threshold to be infinity. We adopt the ratio $f_{\rm HP}^2 \left<\xi^n\right>_{\rm HP}/(f_{\rm HP}^2\left<\xi^0\right>_{\rm HP})$ to derive the $n_{\rm th}$-moment
$\left<\xi^n\right>_{\rm HP}$ instead of directly calculating
$\left<\xi^n\right>_{\rm HP}$. Due to the theoretical uncertainty sources for $f_{\rm HP}$ and $\left<\xi^n\right>_{\rm HP}$ are mutually correlated with each other, such a treatment result in a much smaller theoretical uncertainty. Our results are presented in Table \ref{thpxiBW}, in which the HP leading-twist DA moments $\left<\xi^n\right>_{\rm HP}$ up to $6_{\rm th}$-order are presented. We take the Borel window $M^2 \in [1,2](\textrm{GeV}^2)$ for $\left<\xi^n\right>_{\eta_c}$, $M^2 \in [15,20](\textrm{GeV}^2)$ for $\left<\xi^n\right>_{B_c}$ and $\left<\xi^n\right>_{\eta_b}$, respectively. Fig.\ref{fhpxin} shows the stability of the moments within those allowable Borel windows. In doing the calculation, all the uncertainty sources, such as the Borel parameter, the
dimension-four condensate $\left<\alpha_s G^2\right>$, the dimension-six condensate $\left<g_s^3 f G^3\right>$ and the bound state parameters, have been taken into consideration. The errors listed in Table \ref{thpxiBW} are dominated by varying $M^2$ within the Borel window. The scale $\mu$ is set to be $\bar{m}_c(\bar{m}_c)=1.275{\rm GeV}$ for $\eta_c$ and $\bar{m}_b(\bar{m}_b)=4.18{\rm GeV}$ for $B_c$ and $\eta_b$.

\begin{table}[htb]
\caption{The HP leading-twist DA Gegenbauer moments $a_n^{\rm HP}$ up to $6_{\rm th}$-order, which are derived from $\left<\xi^n\right>_{\rm HP}$ via the relations (\ref{rel_mom1}, \ref{rel_mom2}, \ref{rel_mom3}, \ref{rel_mom4}, \ref{rel_mom5}, \ref{rel_mom6}). The scale $\mu$ is set to be $\bar{m}_c(\bar{m}_c)$ for $\eta_c$ and $\bar{m}_b(\bar{m}_b)$ for $B_c$ and $\eta_b$.}
\begin{tabular}{ c | c c c c }
\hline
~ & ~$\eta_c(\mu = \bar{m}_c(\bar{m}_c))$~ & ~$B_c(\mu = \bar{m}_b(\bar{m}_b))$~ & ~$\eta_b(\mu = \bar{m}_b(\bar{m}_b))$~ \\
\hline
~$a_1$~& ~$      0      $~ & ~$0.466 \pm 0.038$~ & ~$0$~ \\
~$a_2$~& ~$-0.372 \pm 0.027$~ & ~$-0.053 \pm 0.016$~ & ~$-0.387 \pm 0.019$~ \\
~$a_3$~& ~$      0      $~ & ~$-0.106 \pm 0.018$~ & ~$0$~ \\
~$a_4$~& ~$ 0.124 \pm 0.029$~ & ~$-0.028 \pm 0.010$~ & ~$0.136 \pm 0.022$~ \\
~$a_5$~& ~$      0      $~ & ~$-0.017 \pm 0.002$~ & ~$0$~ \\
~$a_6$~& ~$-0.025 \pm 0.017$~ & ~$-0.014 \pm 0.001$~ & ~$-0.028 \pm 0.013$~ \\
\hline
\end{tabular}
\label{thpanmu}
\end{table}

Second, we adopt the relationship between the moments $\left<\xi^n\right>_{\rm HP}$ and the Gegenbauer moments $a_n^{\rm HP}$, i.e. Eqs.(\ref{rel_mom1}, \ref{rel_mom2}, \ref{rel_mom3}, \ref{rel_mom4}, \ref{rel_mom5},  \ref{rel_mom6}), to derive the Gegenbauer moments $a_n^{\rm HP}$ from Table \ref{thpxiBW}. The results for the Gegenbauer moments $a_n^{\rm HP}$ are shown in Table \ref{thpanmu}.

\begin{figure}[tb]
\centering
\includegraphics[width=0.5\textwidth]{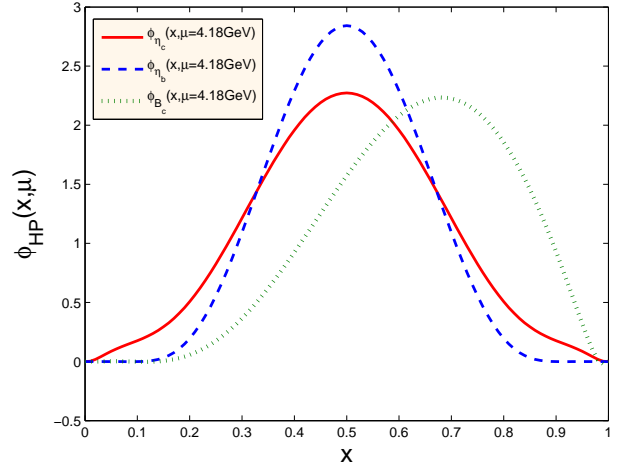}
\caption{The HP leading-twist DAs. The solid, the dotted and the dashed lines are for $\eta_c$ DA, $B_c$ DA and $\eta_b$ DA at the scale $\bar{m}_b(\bar{m}_b)$, respectively.}   \label{fhpphigen}
\end{figure}

Third, we determine all the input parameters $A_{\rm HP}$, $B_n^{\rm HP}$ and $\beta_{\rm HP}$ for the HP leading-twist DA model (\ref{etacDA}). Using the central values for the Gegenbauer moments $a_n^{\rm HP}$ listed in Table \ref{thpanmu}, we obtain, at the scale $\mu = \bar{m}_b(\bar{m}_b)$,
\begin{eqnarray}
A_{\eta_c} &=& 2.401 \textrm{GeV}^{-1}, \nonumber\\
B_2^{\eta_c} &=& -0.306, \nonumber\\
B_4^{\eta_c} &=& 0.092, \nonumber\\
B_6^{\eta_c} &=& -0.019, \nonumber\\
\beta_{\eta_c} &=& 5.386 \textrm{GeV},
\label{parameters}
\end{eqnarray}
for the $\eta_c$ leading-twist DA; and
\begin{eqnarray}
A_{B_c} &=&  1.894 \textrm{GeV}^{-1}, \nonumber\\
B_1^{B_c} &=&  0.400, \nonumber\\
B_2^{B_c} &=&  -0.150, \nonumber\\
B_3^{B_c} &=&  -0.152, \nonumber\\
B_4^{B_c} &=&  -0.014, \nonumber\\
B_5^{B_c} &=&  0.009, \nonumber\\
B_6^{B_c} &=&  -0.001, \nonumber\\
\beta_{B_c} &=&  7.538 \textrm{GeV},
\label{parametersb}
\end{eqnarray}
for the $B_c$ leading-twist DA; and
\begin{eqnarray}
A_{\eta_b} &=& 7.432 \textrm{GeV}^{-1}, \nonumber\\
B_2^{\eta_b} &=& -0.383, \nonumber\\
B_4^{\eta_b} &=& 0.129, \nonumber\\
B_6^{\eta_b} &=& -0.028, \nonumber\\
\beta_{\eta_b} &=& 3.811 \textrm{GeV},
\label{parametersb}
\end{eqnarray}
for the $\eta_b$ leading-twist DA. All those three HPs' leading-twist DAs are presented in Fig.\ref{fhpphigen}. The $\phi_{\eta_c}(x,\mu)$ is broader than $\phi_{\eta_b}(x,\mu)$, and both of them are symmetric, while the $\phi_{B_c}(x,\mu)$ is non-symmetrical, which is consistent with the fact that its constitute $c$- and $b$- quarks are different.

\begin{figure}[tb]
\centering
\includegraphics[width=0.5\textwidth]{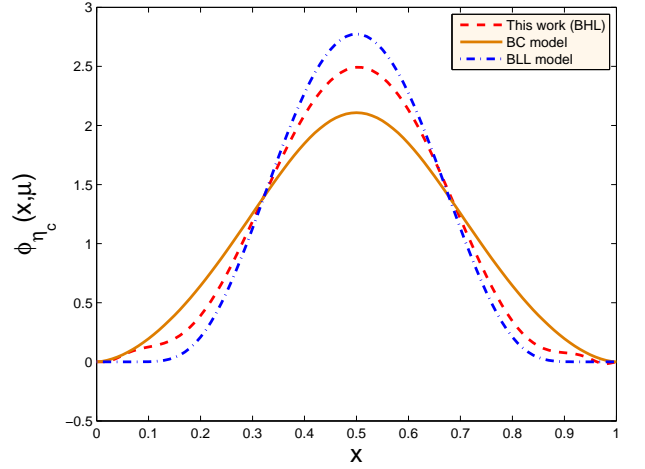}
\caption{A comparison of the $\eta_c$ leading-twist DA. The dashed, the solid and the dash-dot lines are for our present model (\ref{etacDA}), the BC model~\cite{ECDA_BC} and the BLL model~\cite{ECDA_BLL}, respectively. $\mu = \bar{m}_c(\bar{m}_c)$. }
\label{fhpphimodel}
\end{figure}

\begin{figure}[tb]
\centering
\includegraphics[width=0.5\textwidth]{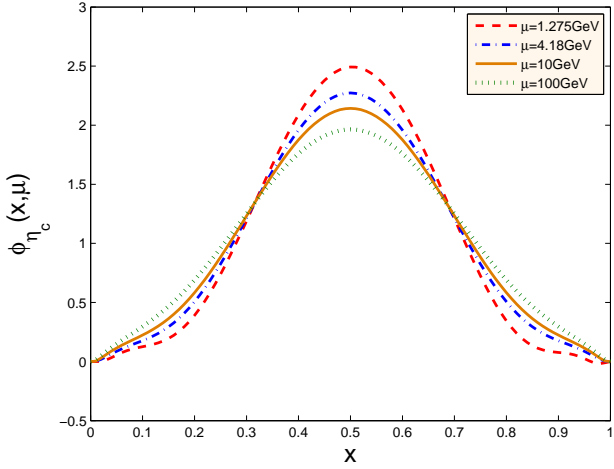}
\caption{The running of the $\eta_c$ leading-twist DA. The dashed, the dash-dot, the solid and the dotted lines are for $\mu = 1.275\textrm{GeV}$, $4.18\textrm{GeV}$, $10\textrm{GeV}$ and $100\textrm{GeV}$, respectively.}
\label{fhpphimu}
\end{figure}

Finally, we take the $\eta_c$ leading-twist DA as an explicit example to show the HP DA properties in detail. Fig.\ref{fhpphimodel} presents a comparison of our $\eta_c$ leading-twist DA model (\ref{etacDA}) with those of the BC model~\cite{ECDA_BC} and the BLL model~\cite{ECDA_BLL}. Our DA model is broader in shape than that of the BLL model, but narrower than that of the BC model. Fig.\ref{fhpphimu} shows how $\phi_{\eta_c}(x,\mu)$ changes with the scale, in which four typical values, i.e. $\mu = 1.275\textrm{GeV}$, $4.18\textrm{GeV}$, $10\textrm{GeV}$ and $100\textrm{GeV}$, are adopted. From Fig.\ref{fhpphimu}, one may observe that with increment of the scale $\mu$, the $\phi_{\eta_c}(x,\mu)$ becomes broader and broader, which shall finally tends to the asymptotic form for $\mu=\infty$ limit.

\subsection{An Application of the Leading-Twist DA $\phi_{\eta_c}$}

As an application, in this subsection, we calculate the $B_c \to \eta_c$ TFF $f_+^{B_c \to \eta_c}(q^2)$ by using our present $\eta_c$ DA model (\ref{etacDA}).

\begin{figure}[htb]
\centering
\includegraphics[width=0.5\textwidth]{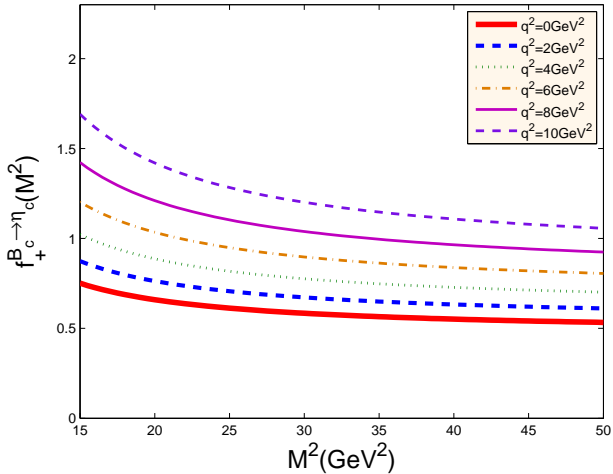}
\caption{The TFF $f_+^{B_c \to \eta_c}(q^2)$ versus the Borel parameter $M^2$ at several typical $q^2$. All the input parameters are taken to be their central values. }
\label{tffm}
\end{figure}

As has been discussed in the Introduction, it is helpful to apply the LCSRs approach with chiral current correlator to calculate $f_+^{B_c \to \eta_c}(q^2)$~\cite{TFF}. Thus the most uncertain twist-3 DAs' contributions are eliminated, and we can see more clearly the properties of the leading-twist DA. Following the standard way as programmed in Ref.~\cite{TFF}, we obtain
\begin{eqnarray}
f_+^{B_c \to \eta_c}(q^2) &=& \frac{\hat{m}_b (\hat{m}_b + \hat{m}_c) f_{\eta_c}}{m_{B_c}^2 f_{B_c}} e^{m_{B_c}^2 /M^2} \int^1_\Delta du \frac{\phi_{\eta_c}(u)}{u} \nonumber\\
&& \times\exp \left[ -\frac{\hat{m}_b^2 - \bar{u} (q^2 - u m_{\eta_c}^2)}{u M^2} \right] \nonumber\\
&& +\textrm{twist-4 and higher-twist terms},
\label{TFF}
\end{eqnarray}
where $\bar{u} = 1-u$, and
\begin{eqnarray}
\Delta &=& \left[ \sqrt{(s_0 - q^2 - m_{\eta_c}^2)^2 + 4 m_{\eta_c}^2 (\hat{m}_b^2 - q^2)} \right. \nonumber\\
&& \left. -(s_0 - q^2 - m_{\eta_c}^2) \right] / (2 m_{\eta_c}^2).
\end{eqnarray}
We take the $\eta_c$ leading-twist DA $\phi_{\eta_c}(u)$ at the scale $\mu \simeq \bar{m}_b(\bar{m}_b)$ to do the calculation. We adopt the same criteria as those of Ref.~\cite{TFF} to determine the Borel window of the process and we take the continuum threshold to be $s_0 = 42 \textrm{GeV}^2$. The determined Borel window is $M^2 = (20 - 35) \textrm{GeV}^2$, in which the TFF also has a good stability as shown by Fig.\ref{tffm}.

\begin{figure}[htb]
\centering
\includegraphics[width=0.5\textwidth]{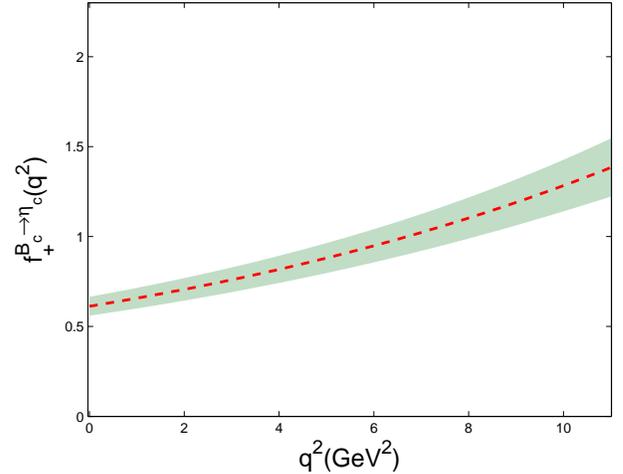}
\caption{The TFF $f_+^{B_c \to \eta_c}(q^2)$ versus $q^2$, in which the shaded hand indicates its uncertainties. }
\label{tffq}
\end{figure}

\begin{table}[htb]
\caption{The fitted parameters $a$ and $b$ for the TFF extrapolation (\ref{extr}). The lowest, middle and the highest TFFs determined from the LCSRs (\ref{TFF}) are adopted for such a determination. }
\begin{tabular}{ c c c }
\hline
~$f_+^{B_c \to \eta_c}(0)$~ & ~$a$~ & ~$b$~ \\
~$0.665$~ & ~$0.072302$~ & ~$0.00040851$~ \\
~$0.612$~ & ~$0.071434$~ & ~$0.00025876$~ \\
~$0.560$~ & ~$0.070434$~ & ~$0.00006892$~ \\
\hline
\end{tabular}
\label{tffpar}
\end{table}

\begin{figure}[htb]
\centering
\includegraphics[width=0.5\textwidth]{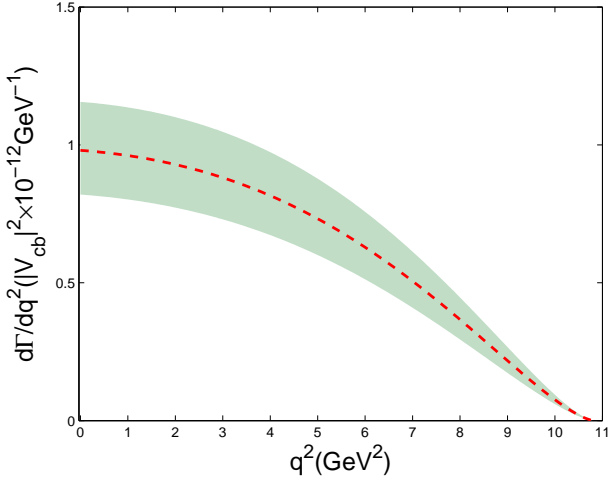}
\caption{The differential decay rate for $B_c \to \eta_c l \nu$ versus $q^2$, where the shaded hand indicates the uncertainty from the TFF $f_+^{B_c \to \eta_c}(q^2)$ only.}
\label{sddrv}
\end{figure}

We present the TFF $f_+^{B_c \to \eta_c}(q^2)$ versus $q^2$ in Fig.\ref{tffq}, in which the shaded hand indicates its uncertainties. At the maximum recoil region with $q^2=0$, we obtain
\begin{eqnarray}
f_+^{B_c \to \eta_c}(0) = 0.612^{+0.053}_{-0.052},   \label{tff0}
\end{eqnarray}
where all uncertainties have been added up in quadrature. Because the LCSRs for the TFF $f_+^{B_c \to \eta_c}(q^2)$ are reliable in low and intermediate regions only, we make use of the following formulae to extrapolate our present prediction to large $q^2$ region~\cite{TFF091,EXTR_TFF},
\begin{eqnarray}
f_+^{B_c \to \eta_c}(q^2) = f_+^{B_c \to \eta_c}(0) \times \exp \left[ a q^2 + b (q^2)^2 \right].
\label{extr}
\end{eqnarray}
The extrapolated high-$q^2$ behavior for the TFF has already been shown in Fig.\ref{tffq} and we put the fitted parameters $a$ and $b$ in Table \ref{tffpar}.

After the extrapolation, we can use the TFF to deal with the exclusive process $B_c \to \eta_c l \nu$. The semileptonic differential decay rate of $B_c \to \eta_c l \nu$ reads
\begin{eqnarray}
&& \frac{d\Gamma}{dq^2} (B_c \to \eta_c l\nu)
\nonumber\\
&=& \frac{G_F^2 |V_{cb}|^2}{192 \pi^3 m_{B_c}^3} \lambda(q^2)^{3/2} \left[ f^{B_c\to\eta_c}_+(q^2) \right]^2,
\label{SDDR}
\end{eqnarray}
where $\lambda(q^2) = (m_{B_c}^2 + m_{\eta_c}^2 - q^2)^2 - 4m_{B_c}^2 m_{\eta_c}^2$, the Fermi constant $G_F = 1.1663787(6) \times 10^{-5} \textrm{GeV}^{-2}$, and the CKM matrix element $|V_{cb}| = 0.0412^{+0.0011}_{-0.0005}$~\cite{PDG}. Fig.\ref{sddrv} shows the differential decay rate of $B_c \to \eta_c l \nu$ versus $q^2$, where the shaded hand indicates the uncertainty from the TFF $f_+^{B_c \to \eta_c}(q^2)$ only. After doing integrating over $q^2\in [0,(m_{B_c} - m_{\eta_c})^2]$, we obtain the central decay width $\Gamma(B_c \to \eta_c l \nu) = 1.12 \times 10^{-14} \textrm{GeV}$. By further using the lifetime of the $B_c$ meson $\tau_{B_c} = (0.452 \pm 0.032) \times 10^{-12} s$~\cite{PDG}, we predict the branching ratio of $B_c \to \eta_c l \nu$ as
\begin{widetext}
\begin{eqnarray}
Br(B_c \to \eta_c l \nu) = (7.70^{+1.50}_{-1.36}|_{f^{B_c\to\eta_c}_+(q^2)}\ ^{+0.42}_{-0.19}|_{|V_{cb}|} \pm 0.02 |_{m_{B_c}} \mp 0.01 |_{m_{\eta_c}} \pm 0.55 |_{\tau_{B_c}}) \times 10^{-3}.
\label{br}
\end{eqnarray}
\end{widetext}
It is noted that the TFF $f^{B_c\to\eta_c}_+(q^2)$, the CKM matrix element $|V_{cb}|$ and the $B_c$ meson lifetime $\tau_{B_c}$ provide the dominant error sources for the branching ratio.

\begin{table}[htb]
\caption{The branching ratio of $B_c \to \eta_c l \nu$ (in unit $\%$). As a comparison, we also present those derived by the LCSRs, the quark model (QM), the pQCD, the QCD relativistic potential model (RPM) and the NRQCD approaches. }
\begin{tabular}{ c c c }
\hline
~Approach~ & ~$Br(B_c \to \eta_c l \nu)$~ & ~Ref.~ \\
\hline
~LCSRs~ & ~$0.770^{+0.165}_{-0.148}$~ & ~This work~ \\
~ & ~$0.75$~ & ~\cite{BRSR03}~ \\
\hline
~QM~ & ~$0.81$~ & ~\cite{FHP061}~ \\
~ & ~$0.48^{+0.02}$~ & ~\cite{FHP062}~ \\
~ & ~$0.67^{+0.11}_{-0.13}$~ & ~\cite{TFF091}~ \\
~ & ~$0.42$~ & ~\cite{TFF03}~ \\
\hline
~pQCD~ & ~$0.441^{+0.122}_{-0.109}$~ & ~\cite{TFF13}~ \\
\hline
~QCD RPM~ & ~$0.15$~ & ~\cite{BRRPM00}~ \\
\hline
~NRQCD~ & ~$2.1^{+0.7}_{-0.3}$~ & ~\cite{BRNRQCD13}~ \\
\hline
\end{tabular}
\label{tbr}
\end{table}
We put our prediction of the branching ratio together with the typical prediction under various approaches in Table \ref{tbr}. It shows that our result agrees with the previous LCSRs estimation~\cite{BRSR03} and also in agreement with the quark model prediction~\cite{FHP061,TFF091} \footnote{A larger branching ratio for $B_c\to\eta_c(J/\psi)l\nu$ is helpful for solving the puzzle for the parameter $\Re(J/\psi \ell^+ \nu)$. A recent discussion on this point can be found in Ref.\cite{shen}.}.

\section{summary}

The meson DA is an important component for the QCD exclusive processes that are studied within the framework of the QCD sum rules, the QCD LCSRs, and the pQCD factorization approaches. The QCD SVZ sum rules provides one of the most effective approaches for exclusive processes, which separates the short- and long-distance quark-gluon interaction, and parameterizes the latter as a series of non-perturebative vacuum condensates. The BFT provides a systematic
method for achieving the goal of SVZ sum rules and also provides a physical picture for the vacuum condensates. As a sequential work of Ref.~\cite{BFSR}, in this paper, we have made a detailed study on the HP leading-twist DAs together with the HP decay constants under the framework of BFT up to dimension-six condensates.

Using the sum rules (\ref{f_srborel}), we obtain $f_{\eta_c} = 453 \pm 4 \textrm{MeV}$, $f_{B_c} = 498 \pm 14 \textrm{MeV}$ and $f_{\eta_b} = 811 \pm 34 \textrm{MeV}$. These values are in agreement with those derived by the Lattice QCD~\cite{FHPL}. Using the sum rules (\ref{srborel}), we calculate the first several moments for the HP leading-twist DA, which are presented in Table \ref{thpxiBW}. Using the relations (\ref{rel_mom1}, \ref{rel_mom2},
\ref{rel_mom3}, \ref{rel_mom4}, \ref{rel_mom5}, \ref{rel_mom6}), we further obtain the Gegenbauer moments up to $6_{\rm th}$-order. More explicitly, the non-zero Gegenbauer moments for $\phi_{\eta_c}$ are: $a_2(\bar{m}_c(\bar{m}_c))=-0.372\pm0.027$,
$a_4(\bar{m}_c(\bar{m}_c))=0.124\pm0.029$ and $a_6(\bar{m}_c(\bar{m}_c))=-0.025\pm0.017$; the non-zero Gegenbauer moments for $\phi_{\eta_b}$ are: $a_2(\bar{m}_b(\bar{m}_b))=-0.387\pm0.019$,
$a_4(\bar{m}_b(\bar{m}_b))=0.136\pm0.022$ and $a_6(\bar{m}_b(\bar{m}_b))=-0.028\pm0.013$; the non-zero Gegenbauer moments for $\phi_{B_c}$ are: $a_1(\bar{m}_b(\bar{m}_b))=0.466\pm0.038$,
$a_2(\bar{m}_b(\bar{m}_b))=-0.053\pm0.016$, $a_3(\bar{m}_b(\bar{m}_b))=-0.106\pm0.018$,
$a_4(\bar{m}_b(\bar{m}_b))=-0.028\pm0.010$,
$a_5(\bar{m}_b(\bar{m}_b))=-0.017\pm0.002$,
$a_6(\bar{m}_b(\bar{m}_b))=-0.014\pm0.001$. Here, the errors are squared average of those from the uncertainties of the Borel parameter, the condensates, and the bound state parameters. The Gegenbauer moments at any other scale can be obtained via evolution.

The meson DA is of non-perturbative nature, thus, it is helpful to have a general model for all the related HPs. Based on the BHL-prescription~\cite{BHL}, we have suggested a model (\ref{etacDA}) for the HP leading-twist DAs. The model parameters of $\phi_{\rm HP}(x,\mu)$ are determined with three reasonable constraints together with the newly obtained HP decay constants and Gegenbauer moments. The behaviors of the $\eta_c$, $B_c$ and $\eta_b$ leading-twist DAs are presented in Fig.\ref{fhpphigen}. It
has been shown that the $\phi_{\eta_c}$ and $\phi_{\eta_b}$ are symmetric and are close in shape; while, the $\phi_{B_c}$ is non-symmetrical and quite different from the naive $\delta$-model, i.e. $\phi_{B_c}(x)\propto \delta(x-\hat{m}_b/m_{B_c})$, suggested in Ref.~\cite{BCDA}. Our present HP DA model can also be adaptable for the light pseudo-scalar DAs, such as pion and kaon DAs. Thus, it shall be applicable for a wide range of QCD exclusive processes. With more and more data available, we may get more definite
conclusions on the behaviors of the pseudo-scalar DAs, and then achieve a more accurate theoretical prediction on those processes.

As an application for the $\eta_c$ leading-twist DA $\phi_{\eta_c}$, we study the TFF $f_+^{B_c \to \eta_c}(q^2)$ within the LCSRs. It is noted that the branching ratio $Br(B_c \to \eta_c l \nu)$ strongly depends on the TFF $f_+^{B_c \to \eta_c}(q^2)$, thus a more accurate TFF shall result in a more accurate branching ratio. At the maximum recoil point, we obtain $f_+^{B_c \to \eta_c}(0) = 0.612^{+0.053}_{-0.052}$. Furthermore, by using the extrapolated
TFF, we predict the branching ratio of the semi-leptonic decay $B_c \to \eta_c l \nu$, i.e., $Br(B_c \to \eta_c l \nu) = 7.70^{+1.65}_{-1.48} \times 10^{-3}$, which is consistent with previous LCSRs prediction~\cite{BRSR03} and the quark model result~\cite{FHP061,TFF091}.

{\bf Acknowledgments}: The authors would like to thank Wen-Fei Wang for helpful discussions. This work was supported in part by the Natural Science Foundation of China under Grants No.11075225, No.11235005, and No.11275280, and by the Fundamental Research Funds for the Central Universities under Grant No.CQDXWL-2012-Z002.

\end{document}